\documentclass[preprint]{aastex}

\usepackage{emulateapj5}
\usepackage{apjfonts}
\usepackage{graphicx}     
\usepackage{hyperref}

\newcommand{\n}{\ensuremath{\mem{n}}}

\newcommand{\p}{\ensuremath{\mem{p}}}

\newcommand{\hedr}{\ensuremath{^{3}\mem{He}}}
\newcommand{\hevi}{\ensuremath{^{4}\mem{He}}}
\newcommand{\lisi}{\ensuremath{^{7}\mem{Li}}}
\newcommand{\besi}{\ensuremath{^{7}\mem{Be}}}

\newcommand{\cdr}{\ensuremath{^{13}\mem{C}}}
\newcommand{\czw}{\ensuremath{^{12}\mem{C}}}

\newcommand{\ndr}{\ensuremath{^{13}\mem{N}}}

\newcommand{\ose}{\ensuremath{^{16}\mem{O}}}

\newcommand{\nadr}{\ensuremath{^{23}\mem{Na}}}

\newcommand{\iac}{\ensuremath{^{128}\mem{I}}}

\newcommand{\msun}{\ensuremath{\, {\rm M}_\odot}}
\newcommand{\lsun}{\ensuremath{\, {\rm L}_\odot}}

\newcommand{\kelv}{\ensuremath{\,\mathrm K}}
\newcommand{\jahre}{\ensuremath{\, \mathrm{yr}}}
\newcommand{\mzams}{\ensuremath{M_{\rm ZAMS}}}
\newcommand{\beq}{\begin{equation}}
\newcommand{\beqa}{\begin{eqnarray}}
\newcommand{\eeq}{\end{equation}}
\newcommand{\eeqa}{\end{eqnarray}}
\newcommand{\bedis}{\begin{displaymath}}
\newcommand{\edis}{\end{displaymath}}

\newcommand{\mem}[1]{\ensuremath{\mathrm{ #1}}}

\newcommand{\punkt}{\mathrm{ \hspace*{0.1cm}.}}

\newcommand{\apleq}{\ensuremath{\stackrel{<}{_\sim}}}
\newcommand{\apgeq}{\ensuremath{\stackrel{>}{_\sim}}}

\newcommand{\kap}[1]{Sect.\,\ref{#1}}
\newcommand{\sect}[1]{Sect.\,\ref{#1}}

\newcommand{\abb}[1]{Fig.\,\ref{#1}}
\newcommand{\fig}[1]{Fig.\,\ref{#1}}

\newcommand{\tab}[1]{Table\,\ref{#1}}

\newcommand{\spr}{\mbox{$s$-process}}
\newcommand{\sprn}{\mbox{$s$ process}}

\slugcomment{LA-UR 10-00630}

\usepackage{natbib}

\shorttitle{Convective-reactive burning in Sakurai's object}
\shortauthors{Herwig etal.}

\begin{document}

\title{Convective-reactive proton-\czw\ combustion in Sakurai's object (V4334 Sagittarii) and implications for the
  evolution and yields from the first generations of stars} 

\author{
Falk Herwig\altaffilmark{1,8}, 
Marco Pignatari\altaffilmark{1,2,3,8}, 
Paul R. Woodward\altaffilmark{4}, 
David H. Porter\altaffilmark{5},
Gabriel Rockefeller\altaffilmark{6,8},
Chris L. Fryer\altaffilmark{6,8},
Michael Bennett\altaffilmark{7,8}, 
and Raphael Hirschi\altaffilmark{7,8,9},
}
\altaffiltext{1}{Department of Physics \& Astronomy, University of Victoria, Victoria, BC V8P5C2, Canada}
\altaffiltext{2}{Joint Institute for Nuclear Astrophysics, University of Notre Dame, Notre Dame, IN 46556, USA}
\altaffiltext{3}{TRIUMF, 4004 Wesbrook Mall, Vancouver, BC V6T2A3, Canada}
\altaffiltext{4}{LCSE \& Department of Astronomy, University of Minnesota, Minneapolis, MN 55455, USA}
\altaffiltext{5}{Minnesota Supercomputing Institute, University of Minnesota, MN, USA}
\altaffiltext{6}{Computational Computer Science Division, Los Alamos National Laboratory, Los Alamos, NM 87545, USA; Physics Department, University of Arizona, Tucson, AZ 85721, USA}
\altaffiltext{7}{Astrophysics group, Keele University, Lennard-Jones Lab., Keele, ST55BG, UK}
\altaffiltext{8}{NuGrid collaboration}
\altaffiltext{9}{Institute for the Physics and Mathematics of the Universe, University of Tokyo, 5-1-5 Kashiwanoha, Kashiwa 277-8583, Japan}

\email{fherwig@uvic.ca}

\begin{abstract}
  Depending on mass and metallicity as well as evolutionary phase,
  stars occasionally experience convective-reactive nucleosynthesis
  episodes. We specifically investigate the situation when
  nucleosynthetically unprocessed, H-rich material is convectively
  mixed with a He-burning zone, for example in convectively unstable
  shell on top of electron-degenerate cores in AGB stars, young white
  dwarfs or X-ray bursting neutron stars. Such episodes are frequently
  encountered in stellar evolution models of stars of extremely low or
  zero metal content, such as the first stars. We have carried out
  detailed nucleosynthesis simulations based on stellar evolution
  models and informed by hydrodynamic simulations. We focus on the
  convective-reactive episode in the very-late thermal pulse star
  Sakurai's object (V4334 Sagittarii).  \citet{asplund:99a} determined
  the abundances of 28 elements, many of which are highly non-solar,
  ranging from H, He and Li all the way to Ba and La, plus the C
  isotopic ratio.  Our simulations show that the mixing evolution
  according to standard, one-dimensional stellar evolution models
  implies neutron densities in the He intershell ($\lesssim$ few
  10$^{11}\mem{cm}^{-3}$) that are too low to obtain a significant
  neutron capture nucleosynthesis on the heavy elements. We have
  carried out 3D hydrodynamic He-shell flash convection simulations in
  $4\pi$ geometry to study the entrainment of H-rich material. Guided
  by these simulations we assume that the ingestion process of H into
  the He-shell convection zone leads only after some delay time to a
  sufficient entropy barrier that splits the convection zone into the
  original one driven by He-burning and a new one driven by the rapid
  burning of ingested H. By making such mixing assumptions that are
  motivated by our hydrodynamic simulations we obtain significantly
  higher neutron densities ($\sim$ few $10^{15}\mem{cm}^{-3}$) and
  reproduce the key observed abundance trends found in Sakurai's
  object. These include an overproduction of Rb, Sr and Y by about 2
  orders of magnitude higher than the overproduction of Ba and
  La. Such a peculiar nucleosynthesis signature is impossible to
  obtain with the mixing predictions in our one-dimensional stellar evolution
  models.  The simulated Li abundance and the isotopic ratio
  $^{12}$C/$^{13}$C are as well in agreement with observations.
  Details of the observed heavy element abundances can be used as a
  sensitive diagnostic tool for the neutron density, for the neutron
  exposure and, in general, for the physics of the convective-reactive
  phases in stellar evolution. For example, the high elemental ratio
  Sc/Ca and the high Sc production indicate high neutron densities.
  The diagnostic value of such abundance markers depends on uncertain
  nuclear physics input. We determine how our results depend on
  uncertainties of nuclear reaction rates, for example for the
  $\cdr(\alpha,\n)\ose$ reaction.
\end{abstract}


\section{Introduction}

\subsection{Convective-reactive phases of stellar evolution}
\label{sec:intro_conv}
In stellar evolution the nuclear time scale is usually much larger
than the convective mixing time scale. However, this is not always the
case. An example of stellar nucleosynthesis where nuclear reactions
and convective mixing occurs on the same time scale are slow neutron
capture process branchings \citep[$s$
  process,][]{burbidge:57,wallerstein:97} in He-shell flash convection
of Asymptotic Giant Branch (AGB) stars, such as the branching at
\iac\ \citep{reifarth:04}. This situation is comparatively simple to
simulate as the rapid nuclear reaction in question, the double-decay
of \iac, does not release any significant amount of energy. A
post-processing approach of the standard stellar evolution calculation
with some one-dimensional treatment of convection, like mixing-length
theory (MLT), with time-dependent mixing gives a reasonable
approximation of this situation.\footnote{Although even in this case
  multi-dimensional effects of convection have to be taken into
  account eventually as simulations by \citet{herwig:06a} indicate
  that the velocity profile at the bottom of the convective shell is
  flatter compared to the MLT prediction.}

The goal of this paper is instead to investigate the situation when
rapid nuclear reactions are indeed releasing amounts of energy that are likely
to affect the fluid flow, as for example in the case of proton capture
of \czw\ in convective He-burning. In the fluid dynamics community
this mixing regime is sometimes refered to as level-3 mixing, where
the flow is altered by attendant changes in the fluid
\citep{dimotakis:05}.  We refer to these situations as
reactive-convective phases in order to emphasize the fact that the
time scales of highly exothermic nuclear reaction and the convective
fluid flow time scales are of the same order. 

The ratio of the mixing time scale and the reaction time scale is
called the Damk\"ohler number: \beq D_\alpha =
\frac{\tau_\mem{mix}}{\tau_\mem{react}} \punkt \eeq MLT is concerned
with averaged properties both in time over many convective turn-overs
and in space over the order of a pressure scale height. In the
categories of \citet{dimotakis:05} diffusion coeffiecients derived
from MLT may describe level-1 mixing (while mixing induced by rotation
involves flow dynamics that are altered by mixing processes and
labeled in this scheme as level-2 mixing). Therefore, time-dependent
mixing through a diffusion algorithm with diffision coefficients
derived from MLT is appropriate for regimes with $D_\alpha \ll 1$.
The difficulty of simulating convective-reactive phases in present
one-dimensional stellar evolution codes then appears as the inability
of MLT (or any similar convection theory) to properly account for the
additional dynamic effects introduced through rapid and dynamically
relevant nuclear energy release in level-3 mixing associated with 
Damk\"ohler numbers $D_\alpha \approx 1$.

Convective-reactive episodes can be encountered in numerous phases of
stellar evolution, including the He-shell flash of AGB stars of
extremely low metal content \citep[e.g.][]{fujimoto:00,suda:04,iwamoto:04,cristallo:09b},
metallicity low-mass stars
\citep[e.g.][]{hollowell:90b,schlattl:02,campbell:08}, young white
dwarfs of solar metallicity
\citep[e.g.][]{iben:83a,herwig:99c,lawlor:03}, both rotating and
non-rotating Pop III massive stars \citep{ekstroem:08}, and more in
general, in low metallicity massive stars \citep{woosley:95}. These
combustion events are encountered as well in X-ray burst calculations
of accreting neutron stars \citep{woosley:04,piro:07}, and accreting
white dwarfs \citep{cassisi:98} that may be the progenitors of SN
Ia. Convective-reactive events have been found in post-RGB stellar
evolution models and associated with the horizontal branch anomalies
in certain globular clusters \citep{brown:01,miller-bertolami:08}.
Finally, again in AGB stars, convective-reactive phases can be found in
hot dredge-up \citep{herwig:03c,goriely:04,woodward:08a}, a phenomenon
that is associated with the treatment of convective boundaries,
generally in more massive and lower metallicity AGB stars.

Although convective-reactive phases are quite common in stellar
evolution, in particular in the early, low-metellicity Universe, we do
not currently have a reliable and accurate way of simulating them.  In
this work we discuss the case of the He-shell flash with H-ingestion
in a very-late (post-AGB) thermal pulse at solar metallicity.  This
situation is extremely similar to H-ingestion associated with the
He-shell flash in AGB stars at extremely low metallicity.  The
one-dimensional, spherically-symmetric stellar evolution approximation
is not very realistic in this case, because both the entrainment of H
into the He-shell flash convection zone as well as the subsequent
convective transport, mixing and nuclear burning of hydrogen enriched
fluid parcels are inherently a three-dimensional hydrodynamic
process. The energy from proton captures by \czw\ via the $\czw(\p,\gamma)\ndr$ reactions is
released on the same time scale ($\sim1 \dots 10 \mem{min}$ for
$T=1.3 \dots 1.05 \times10^8\kelv$) of the fluid flow of convection (\sect{sec:timescale}), and this
energy will add entropy to fluid elements and in turn feedback into
the hydrodynamics \citep{herwig:01a}.  These highly coupled,
multi-dimensional processes are approximated in through the MLT,
complemented with a time-dependent mixing algorithm. This assumption
may not be realistic in the present case (see \sect{sec:nuc1} and
\ref{sec:hydro-implication}).

\subsection{Post-AGB flash star Sakurai's object and its observed abundance properties} 
\label{sec:intro-sak}
Sakurai's object is a very-late thermal pulse
post-AGB object \citep[][and ref.\ there]{duerbeck:00} and has
experienced a H-ingestion flash in 1994. The star's observed abundance
signatures are highly non-solar, and very unusual for a post-AGB
low-mass star (\sect{sec:nuc1}).  Nevertheless, there is wide
agreement in the literature that the object's distance is $2 -
5\mem{kpc}$ and that it has a mass of around $0.6\msun$ \citep[][and
  ref. therein]{vanhoof:07}, pointing to a low mass star progenitor.
Moreover, the high abundance of Li requires the existence of \hedr\ in
the envelope \citep{herwig:00f}, pointing again to a low mass star
progenitor that was not affected by hot bottom burning (HBB). Indeed,
hot-bottom burning occurs at solar metallicity for stars with
$\mzams\apgeq4\msun$ and destroys \hedr\ in the AGB envelope
\citep{scalo:75}. Another process, that could effect the evolution of
\hedr\ during the progenitor evolution of Sakurai's object is
extra-mixing below the convective envelope during either the RGB or
AGB
\citep[e.g.\ ][]{wasserburg:95,charbonnel:07,denissenkov:10a}. Sakurai's
object serves potentially as an important constraint for theories of
such mixing because the observed Li abundance increase during the observations in 1996 as reported by \citet{asplund:99a} can only be
modeled in the very late thermal pulse if significant amounts of
\hedr\ are still present in the envelope at the beginning of the
post-AGB evolution.

The light curve of this object was closely monitored as it evolved
within approximately $2 \jahre$ from the pre-WD location in the HRD
back to the AGB location, a much shorter evolution time scale than
previously predicted \citep{herwig:99c}.  A possible explanation of
such a fast born-again evolution of Sakurai's object is that the
convective mixing efficiency in the He-shell flash convection zone is
smaller by a factor of $\sim 30$ compared to the MLT predictions in standard
one-dimensional stellar models \citep{herwig:01a}. This modification
is motivated by the reasoning that in the convective-reactive regime
the fluid flow would be eventually strongly affected by the energy
released rapidly on a time scale comparable to the fluid flow
velocity.  This process, indeed, would locally add boyuancy to the
fluid element causing a behavior that is not reflected in the
mixing-length theory.

\citet{miller-bertolami:06} have presented a more detailed
investigation and emphasize the importance of appropriate time
resolution. In addition, they studied the role of overshooting and
$\mu$-gradients.  Their simulations with exponential,
depth-dependent overshooting agree better with observations than
tracks computed without any overshooting.  $\mu$-gradients appear to
have only secondary effects.  Confirming the mass dependence of the
proton-ingestion born-again evolution first reported by
\citet{herwig:01a}, \citet{miller-bertolami:07} point out that the
initial return light-curve of Sakurai's object could be reproduced
with a slightly lower mass model than the $0.604\msun$ adopted by Herwig (2001),
a high time resolution and their
alternative description of convective transport. However, the second
heating phase into which the Sakurai's object has entered now 
\citep[][]{vanhoof:07},
seems to be better in agreement with the
modified convection models proposed by \citet{herwig:01a}.

While the light curve of Sakurai's object has certainly raised doubts
about the capability of one-dimensional stellar evolution calculations
to reproduce its evolution, in this work we show that the abundance
determinations by \citet{asplund:99a} pose a much more stringent
constraint on the physics of convective-reactive phases. Asplund et
al.\ determined 28 elemental abundances at four times between April
and October 1996, when the star had cooled to below $8000\kelv$.  
In particular, 
among light elements a significant enhancement (at least 0.5 dex) with respect
to the solar abundance has been observed for Li, Ne and P.
Beyond iron, Cu, Zn, Rb and Sr peak elements are
significantly enhanced. In addition, there are trends as a function
of time that are smaller than the differences to solar.
However, for this initial analysis which is not yet based on full hydrodynamic
simulations with nuclear burn, we will not discuss those trends in detail.

\begin{figure}[tbp]
   \includegraphics[width=0.9\textwidth]{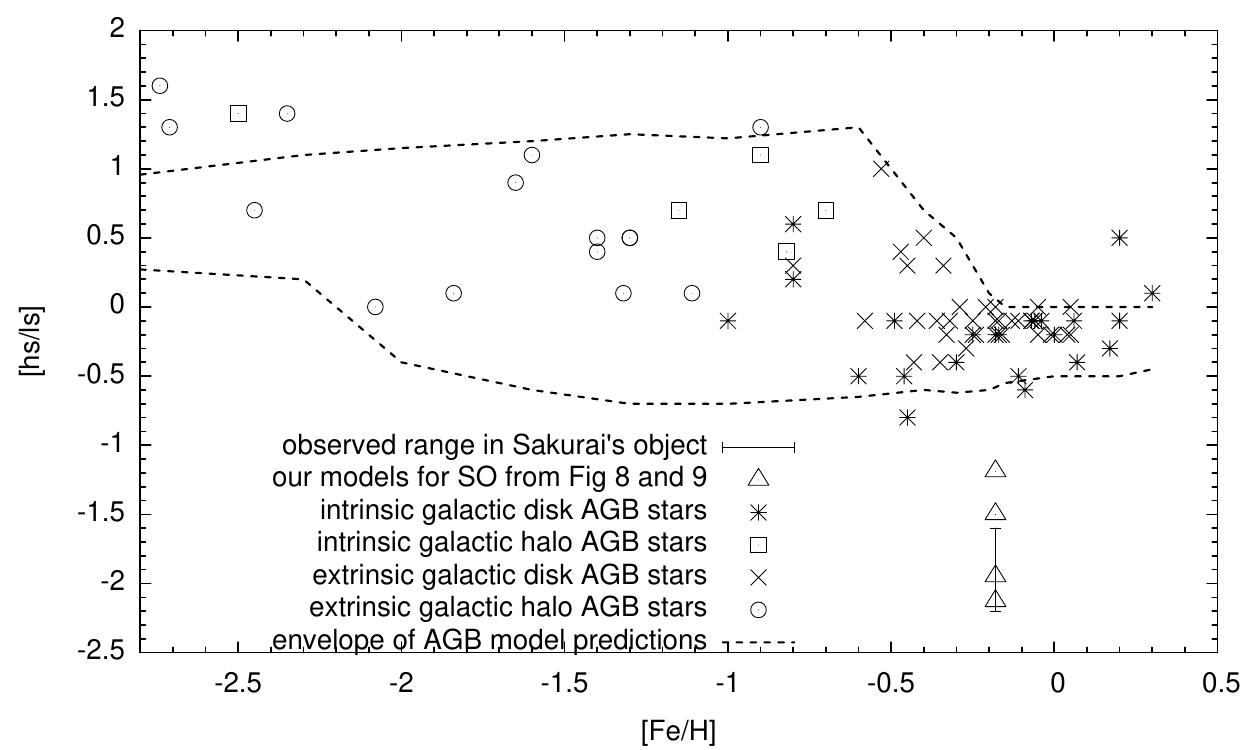}
   \caption{Observed and predicted s-process abundance distribution
     index ratio [hs/ls] for stars with a large range of
     metallicities. Observations
     \citep{tech:71,smith:84,smith:84b,smith:85,smith:86,smith:87b,smith:90,smith:93,smith:96,smith:97,abia:98,vanwinckel:00,zacs:95,zacs:98,zacs:00,reddy:99,kipper:96,kipper:94,tomkin:83,tomkin:86,kovacs:85,vanture:92b,vanture:00,pereira:98,aoki:00,mcwilliam:95,mcwilliam:98,norris:97a,beveridge:94}
     and model predictions of AGB stars are from \citet{busso:01a}. 
     In the Figure, the
     [hs/ls] ratio observations of the Sakurai's object have a certain
     range, depending on which of the 4 observations from Asplund
     et al.\ are considered, and how the indices are calculated. In general,
     the ratio is about $2 \mem{dex}$ smaller compared to AGB
     predictions and observations. 
     Our nucleosynthesis results are also included for comparison 
     (see \sect{sec:nuc2} for details). 
}
   \label{fig:hsls}
\end{figure}

A few preliminary comments on individual elements may be in order.
The observed Li is clearly produced above the meteoritic value. 
\citet{herwig:00f} proposed that together with
protons \hedr\ is ingested into the He-shell flash convection zone, 
providing the fuel to produce Li via the reaction chain
$^{3}$He($\alpha$,$\gamma$)$^{7}$Be($\beta^+$)$^{7}$Li.
The first $s$-process peak elements are
enhanced by up to 2dex while Ba and La are not enhanced, causing a
ratio of Ba peak to Sr peak elements that is much lower than expected
from models and observations of AGB stars \citep{busso:01a}. We can
translate the abundances observed by \citet{asplund:99a} into the
ratio of the two \spr\ indicator indices $hs$ and $ls$. An
s-process index $s/s_{\odot}$ is the overproduction factor of a group
of \spr\ elements with respect to the initial solar value. The index
ratio [hs/ls] = [hs/Fe] - [ls/Fe] monitors the distribution of the
\spr\ elements, and it is an intrinsic index of the neutron capture 
nucleosynthesis on heavy elements \citep{luck:91}.
We have used
[ls/Fe]=$\frac{1}{3}$([Sr/Fe]+[Y/Fe]+[Zr/Fe]) and
[hs/Fe]=$\frac{1}{2}$([Ba/Fe]+[La/Fe]), where square brackets indicate
the logarithmic ratio with respect to the solar ratio
(\tab{tab:asplund_abundances}).  For Asplund's October measurements
the indices are [hs/Fe]$ = 0.05$ and [ls/Fe]$ = 1.9$ assuming that
[Fe/H]$ = 0.0$ for Sakurai's object. 
We record measurements of $\pm 0.2 \dots 0.3 \mem{dex}$ as the average
approximate index ratio [hs/ls]$ \sim -2$ at the end of the observed
period. 
In \fig{fig:hsls}, we compare such ratio with $s$-process theoretical
predictions and stellar observations of low mass AGB stars, that are 
the progenitor population of the Sakurai's object.
In particular, we show that the observed [hs/ls] is a factor of ten or more
lower than in typical AGB stars.
Therefore, the nucleosynthesis environment that has generated the abundances
observed by Asplund et al. was very different from that encountered 
in the previous AGB phase.
In \fig{fig:hsls} we also include [hs/ls] from our nucleosynthesis
calculations presented in this paper, that succesfully reproduce the
same ratio measured in the Sakurai's object.
Such calculations will be discussed in detail 
in Section \ref{sec:nuc2}.

The abundance pattern of Sakurai's object further distinguishes
itself from the AGB stars through the significantly enhanced P, Cu and
Zn. These elements are not usually produced in low-mass
stars. Several other elements are reduced, i.e., S, Ti, Cr and Fe. 
In particular, Fe is expected to be depleted, since it is
the seed for n-capture nucleosynthesis.
All these abundance signatures appear to be the result of a n-capture
burst of large n-density.
Another important feature is the C isotopic ratio $\czw/\cdr \sim 3-4$,
where the large \cdr\ abundance results from the 
$\czw(\p,\gamma)^{13}$N($\beta^+$)$^{13}$C reaction channel.
$^{13}$C is also the main neutron source during the 
H ingestion event, which causes the peculiar abundance signature
observed by Asplund et al. (see Section \ref{sec:nuc2} for details). 

In the following we will briefly describe the tools we use in this
investigation (\sect{sec:codesI}) and defer more details to an
appendix (\sect{sec:codes}). Next we describe the stellar evolution picture of Sakurai's
object and show how nucleosynthesis simulations based directly on the
output of one-dimensional stellar evolution calculations fail to account for the
observed abundance patterns (\sect{sec:1D}). Then we describe
hydrodynamic simulations of entrainment into He-shell flash convection
that motivate our modified mixing assumptions
(\sect{sec:hydro}). We will show how corresponding  nucleosynthesis
simulations account
for the observed abundances, and we discuss the incfluence of nuclear
reaction rate uncertainty (\sect{sec:nuc2}). The paper ends with a
summary and some remarks on implications for the nucleosynthesis in
the first generations of stars, including the light-element primary
process (\sect{sec:disc}). In the appendix we give additional
information on the codes we have used, and in Appendix \ref{sec:timescale} we
discuss time scales for burning and mixing.

\section{Simulation codes}
\label{sec:codesI}
Three different types of simulation codes have been used in this work:
\begin{itemize}
\item a stellar evolution code (EVOL), providing one-dimensional
  stellar evolution up to the post-AGBn and thermodynamic structures
  for the beginning of the post-AGB He-shell flash event, also known
  as the very late thermal pulse (VLTP);
\item a multi-zone post-processing nucleosynthesis code (PPN) with
  complete nuclear network and mixing;
\item  a multi-dimensional-hydrodynamical code (PPM), to study how hydrogen
is ingested during the VLTP.
\end{itemize}

We have used the stellar evolution code EVOL to calculate the global
evolution of post-AGB stars (\sect{sec:stellar_evol}) experiencing a
VLTP \citep{bloecker:95a,herwig:99a,herwig:04b}. The assumptions and
input physics choices are very similara to those in
\citet{herwig:01a}.  Furthermore, we have used structures from the
last thermal pulse of the AGB model by \cite{herwig:04b}, and of the
VLTP model by \cite{herwig:99c}.

For the detailed nucleosynthesis simulations (\sect{sec:nuc1} and
\ref{sec:nuc2}) we have used the PPN (Post-Processing Nucleosynthesis)
code \citep{herwig:08a}. This code allows to calculate the complete
nucleosynthesis along the radial profile of a star according to the
structure input from a stellar evolution model in as many zones as
required. Nuclear burn steps are alternated with time-dependent mixing
steps. Details, including the nuclear physics data information, are given in \kap{sec:mppnp}.

In order to investigate the hydrodynamic behaviour of unprocessed
H-rich material entrained into the He-shell flash convection
(\sect{sec:hydro}), we used Woodward's PPM gas dynamics code with the
PPB advection scheme on a cartesian grid
\citep{woodward:03,woodward:06a,woodward:08c}. For important code
details, see \kap{sec:code-hydro}.

\section{The stellar evolution picture}
\label{sec:1D}
\subsection{Global stellar evolution scenario and calculation}
\label{sec:stellar_evol}
The VLTP evolution scenario involves a He-shell flash
on a single young white dwarf after the end of H-shell burning when
the evolution track has just entered the white dwarf cooling curve in
the HRD, as for example shown in \citet{herwig:99c}, and in more
detail in Sect.\,3.2.1 of \citet{miller-bertolami:06}. It involves the
convective ingestion of all or parts of the small ($\sim
10^{-4}\msun$) remaining unprocessed, and thus H-rich, envelope into
the hot ($T=1 \dots 3 \times 10^8 \kelv$) He-burning flash
layers. This He-burning convection zone contains a mass fraction of $20
\dots 40\%$ \citep[depending on convective model
  assumptions,][]{herwig:99a,miller-bertolami:06} of primary
\czw. Protons are rapidly captured by the abundant \czw, on the time
scale of convective fluid flows of approximately $5\dots10 \mem{min}$.

The progenitor is a low mass AGB star for which \spr\ element
enhancements are expected at the Sr-Y-Zr peak and at the Ba peak
\citep[e.g.][]{busso:01a}.  
elements signature observed in Sakurai's object is not typical of the
\sprn\ in AGB stars.  Indeed, according to the observations by
\citet{asplund:99a}, the ratio of the second peak to the first peak
\spr\ elements is $\mathrm{[Ba/Y]} \sim -2$, in contrast to the
expected AGB stars ratio $-1 < \mathrm{[hs/ls]} < 1$ at solar-like
metallicity \citep[e.g.][]{busso:01a}.  This result does not change if
we assume a lower than solar metallicity for Sakurai's object of
[Fe/H] = -0.63 (values between brackets in
Tab.\ref{hsls:sakurai}). Such a choice may be indicated by the
sub-solar observed Ba abundance, and indeed, the Ba and La abundance
even lead us to assume that there was no significant $s$-process
contribution in the previous AGB phase at all.

In any case, the peculiar abundance signatures of Sakurai's object has
to originate in the H-ingestion event of the VLTP, and can not be
explained in terms of any nucleosynthesis during the AGB progenitor
evolution.

The initial abundance distribution for our post-AGB He-shell flash
nucleosynthesis simulations is a combination of light elements (with
$\mem{A}<23$) from the intershell abundance of an AGB star at the end
of the evolution taken from a 2\msun\ simulation similar to those in
\citet{herwig:04b}, and heavier species according to \cite{asplund:05}
with the isotopic ratios from \citet{lodders:03} scaled to metallicity
$\mem{[Fe/H]} = -0.18$.

The intershell abundances that matter for our
simulations are mostly primary He-burning products, so details of the
initial abundance are not important. 
The choice of more recent solar abundances \citep[][]{asplund:09,lodders:09}
would not modify the results presented in this paper.

In the following section we will discuss the nucleosynthesis according
to one-dimensional stellar evolution mixing preditions of the very-late thermal pulse.

\subsection{Nucleosynthesis according to the stellar evolution model}
\label{sec:nuc1}
\begin{figure}[tbp]
\centering
   \includegraphics[width=0.7\textwidth]{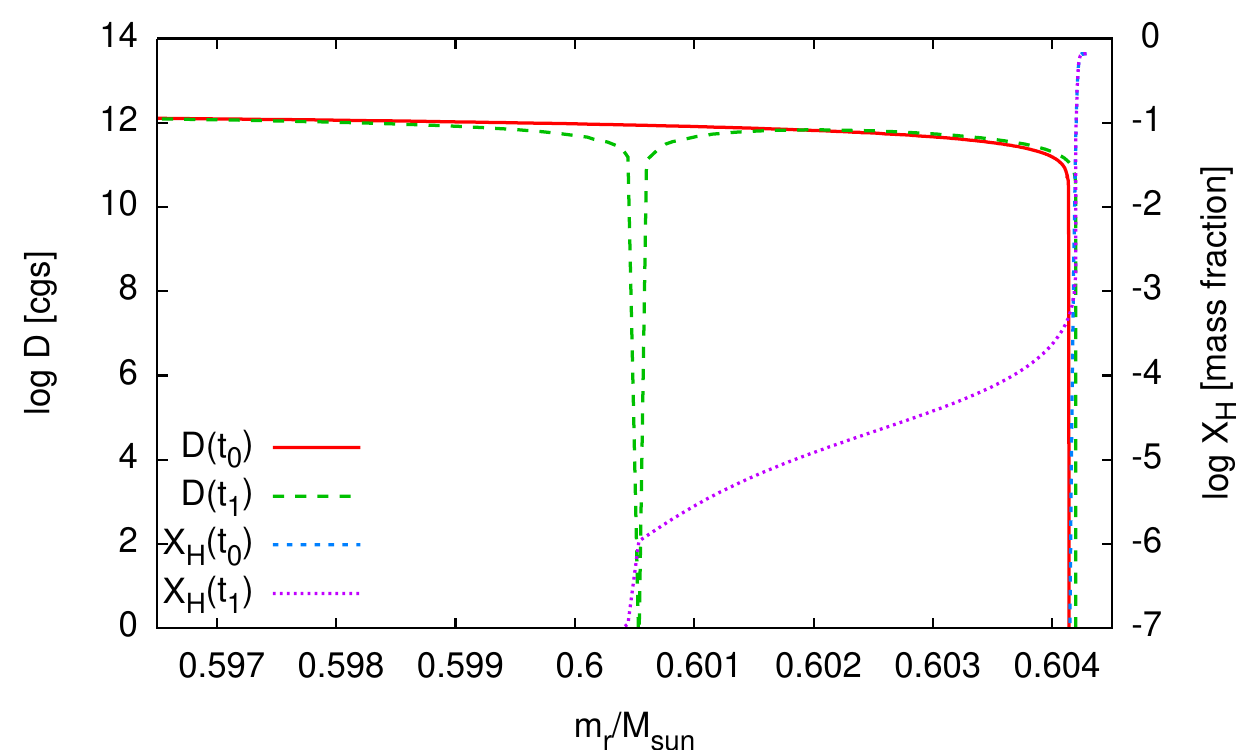}
   \includegraphics[width=0.7\textwidth]{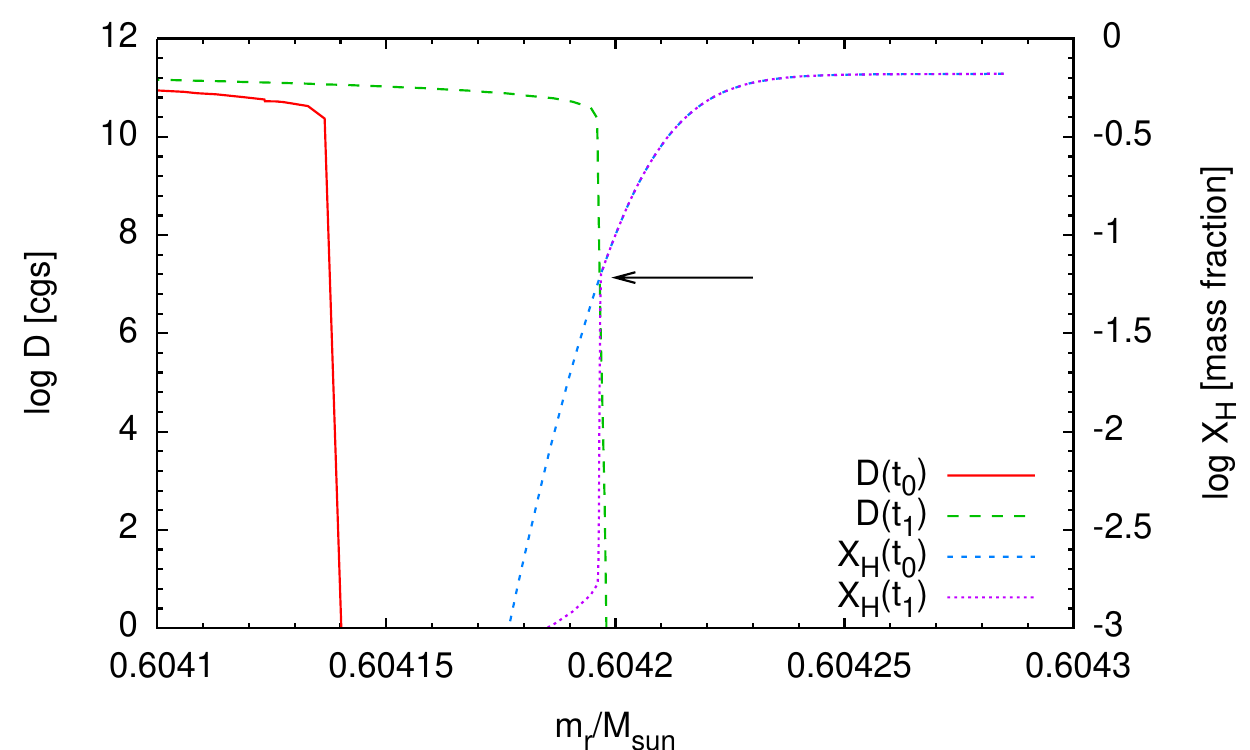}
   \caption{ Convective diffusion coefficient and H abundance profile
     at the beginning of the H-ingestion flash $t_\mem{0}$ and at the
     time when the split of the convection zone appears at
     $t_\mem{1}=t_\mem{0} + 8.58 \cdot 10^{5}\mem{s}$. Top panel: the
     outer section of the convection zone showing the location of the
     split as a deep dip in $D$; bottom panel: just the interface of
     the outer boundary of the convection zone. The arrow indicates
     the H abundance at the position that has been reached by the
     convection zone at the time $t_\mem{1}$. $t_\mem{0}$ is at the
     time of the minimum of the H-burning luminosity at the onset of
     the H-ingestion event.}
   \label{fig:SEprofiles}
\end{figure}

 \fig{fig:SEprofiles} shows the H-profile from stellar evolution in
 the initial phase of the H-ingestion phase for a model like those in
 \citep{herwig:01a}, recalculated with $f_\mem{v}=30$ and higher time resolution. The proton abundance
 at any location is the result of mixing and simultaneous burning. The
 two times correspond to panel A and B in Fig.\ 4 in
 \citet{miller-bertolami:06} and the account of events given in their
 Sect.\,3.2.1 applies here as well.

At time $t_0$ the He-shell flash convection zone is about to make contact
with the H-rich layers above. The H-profile at $m_\mem{r}\sim0.6042\msun$
is the burning profile of the now extinct H-shell. During the late
phase of the post-AGB evolution, basically past the 'knee' in the HRD,
the H-shell is inactive, and the He-shell convection can grow into the
H-rich layers and mix those protons (and \hedr) down into the
\czw-rich He-shell flash convection zone. As H is mixed into deeper
and hotter regions its lifetime against capture by \czw\ decreases
because the rate of the nuclear reaction $\czw(\p,\gamma)\ndr$ increases
strongly with temperature. At some depth, in our simulation at
$m_\mem{r}=0.6005\msun$, the mixing time scale equals the nuclear time
scale (Damk\"ohler number $Da \sim 1$, \sect{sec:intro_conv}) and protons
are now reacting rapidly with \czw, thereby releasing for a brief
period more energy than the He-shell that is intially driving the
flash. 

In the stellar evolution simulation we treat time-dependent mixing
mathematically as a diffusion process. It is implicitly assumed that
on spheres the H-abundance is exactly homogeneous, and that the
radial mixing efficiency based on the radial mean convective velocity
is also exactly homogeneous. This assumption in combination with the
strong temperature sensitivity of the p-capture reaction causes the
stellar evolution code to predict the shell of peak H-burning energy
release to be extremely thin. In the stellar evolution code an entropy
step develops that separates the H-ingestion top convection from the
He-shell flash convection underneath. A thin radiative zone formally
prohibits mixing between the two convection zones. It shows up as a
break in the diffusion coefficient line for time $t_1$ in the top
panel of \fig{fig:SEprofiles}. It now depends on the convective
boundary mixing assumptions whether or not material from the top
convection zone can mix below and vice versa. These boundary mixing
assumptions, i.e.\ the amount of overshooting appropriate for this
situation, is not yet known.

\fig{fig:SEprofiles} shows that the split of the two convection zones
appears already very early when only a small amount of protons has
been consumed. We mark the position in the H-profile and the
corresponding H-abundance that has been reached at the time when the
split occurs in the lower panel. The good agreement of our evolution
simulation with the result by \citet[][
Fig.\,3 in their work]{miller-bertolami:06} only means that these
calculations properly converge and are precise, 
but not that they are accurate.  


At the time of the split the peak temperature in the now separated top
H-burning driven convection zone is $T \apleq 1.0 \times
10^{8}\kelv$. Although the $\czw(\p,\gamma)\ndr(\beta^+)\cdr$ reaction
chain is providing plenty of the neutron source isotope \cdr, the
$\cdr(\alpha,\n)\ose$ reaction activation depends on the peak
temperature reached in this top convection layer. For $T=10^8\kelv$
the lifetime of \cdr\ against capture by \hevi\ (and thus the
time-scale of releasing neutrons) is $ 454\jahre$, and thus 
\emph{neutron capture nucleosynthesis is negligible}, considering that the
born-again life time is only a few years. As a result,
these stellar evolution models cannot provide the environment
to generate abundance patterns as observed by \citet{asplund:99a}.

We have performed a full nucleosynthesis analysis of the stellar
evolution model sequence shown in \fig{fig:SEprofiles}, using the
MPPNP code (\kap{sec:mppnp}). The technique for this nucleosynthesis
analysis is explained in full detail in \sect{sec:nuc2}. Indeed no
modification of heavy element abundances is seen, in disagreement with
the observations by \citet{asplund:99a}, and in agreement with the
qualitative arguments that these authors made in their original paper.

The \citet{herwig:99c} models show a larger peak-temperature of
$T=1.5\times10^8\kelv$\footnote{We have now recalculated those old models
  with higher resolution and find the peak H-burning location at
  slightly lower temperature of $T=1.3\times10^8\kelv$.} for the H-ingestion driven top convection
zone. As discussed in detail in \citet{herwig:01a}, those older models
are not correctly reproducing the fast luminosity rise time observed
in Sakurai's object,
and it exists an
inverse correlation between the rise time and the depth of the burning
zone and split 
(i.e. convection speed, peak temperature). 
models with the higher peak temperature have far too slow rise times
and can thus be excluded. For these higher peak temperatures the life
time of \cdr\ is $0.13\jahre$. However, even this is not short enough
to generate the abundance patterns observed in Sakurai's object (see
\sect{sec:nuc2} for further discussion).

We conclude from this analysis that a one-dimensional stellar
evolution calculation cannot fully account for the mixing conditions
in the convective-reactive H-ingestion flash that occured in Sakurai's
object. In this section we have already hinted at the possible reasons
for the decrepancy. We will now have a closer look at what information
and guidance we can derive from present three-dimensional
hydrodynamic simulations of He-shell flash convection.

\section{The hydrodynamic picture}
\label{sec:hydro}

\subsection{New simulations}
\label{sec:sim}
\begin{figure}[tbp]
   \includegraphics[width=0.503\textwidth]{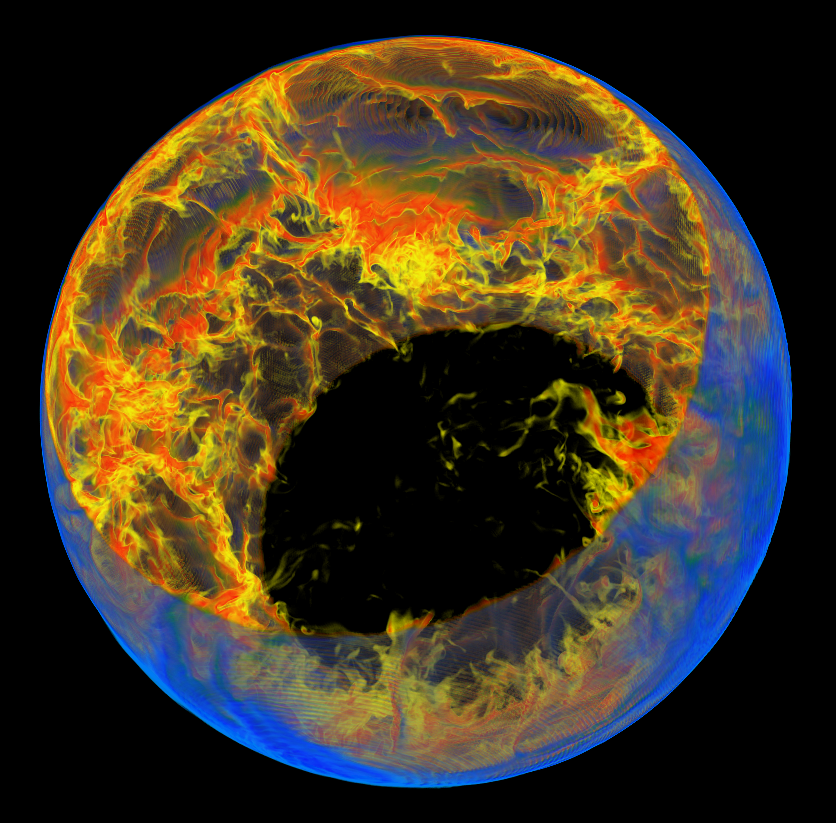} 
   \includegraphics[width=0.497\textwidth]{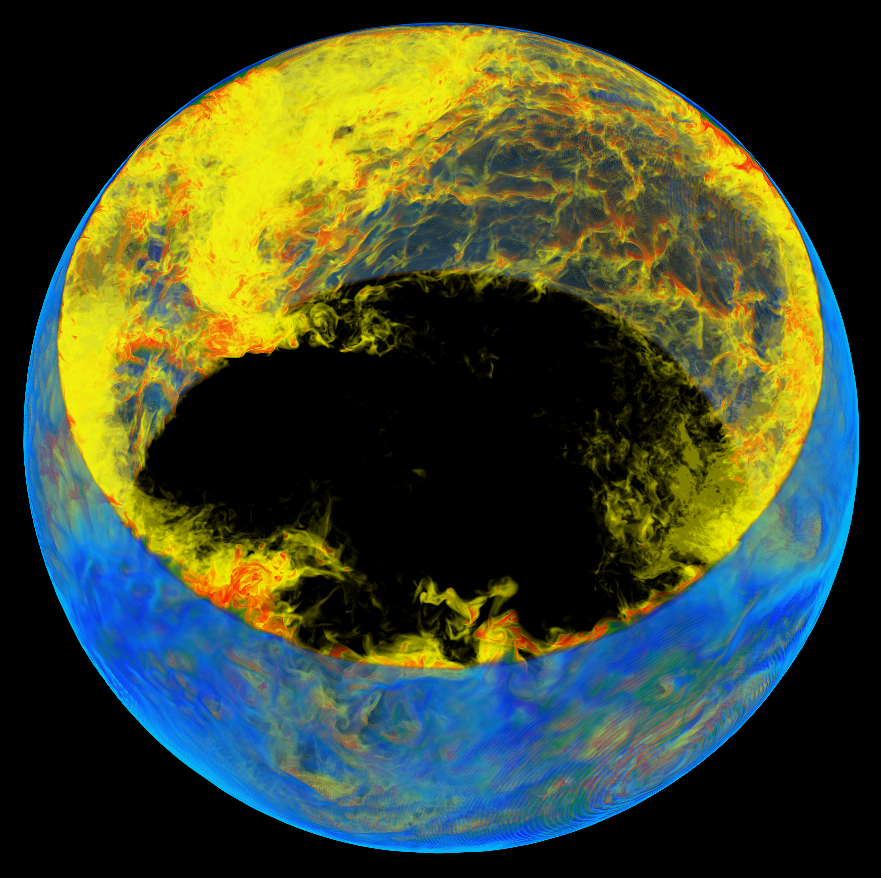} 
   \caption{ Hydrodynamic picture of H-entrainment into He-shell flash
     convection near the luminosity peak of the flash. The setup is based
     on a stellar evolution model corresponding to the situation
     shortly after time $t_\mem{0}$ shown in \fig{fig:SEprofiles},
     when the top of the convection zone is just making contact
     with the H-rich stable layer.  Colors indicate abundance of
     proton-rich material that is originally only in the stable layer
     above the convection zone that is entrained into the convection
     zone. Volume fractions of about $\sim 1\%$ are shown as blue, while
     concentrations that are close to one are transparent. The lowest
     concentration yellow blobs that are mixed deep into the
     convection zone correspond to $\sim 0.01\%$. Abundance levels below
     approximately $5\times10^{-5}$ have been made transparent as
     well. The left panel shows a snapshot from a $384^3$ grid while
     the right panel image is from a run on a $576^3$ grid. Slightly
     different times are shown and similar but not identical color
     maps have been used. The PPM simulation is described in more
     detail in \sect{sec:sim}, and the simulation code is described in
     \sect{sec:code-hydro}.  }
   \label{fig:3Dfluids}
\end{figure}

In order to study the hydrodynamic process of entrainment and further
mixing of H-rich material from the stable layers into the convection
zone we have carried out new gas dynamics simulations of the entire
three-dimensional He-shell flash convection domain in $4\pi$ geometry
(\fig{fig:3Dfluids}). We used the PPM code described in
\sect{sec:code-hydro}. We have not included burning of protons with
\czw\ because we restrict the goal of the numerical experiments purely
to the investigation of mixing properties during the onset of the
H-ingestion, which starts when the He-shell flash convecion has
reached its largest Lagrangian extension.

 \citet{herwig:06a}
simulated the He-shell flash convection shell as plane-parallel
box-in-a-star. They selected an earlier phase of the He-shell flash
when the convection had not yet reached its largest extent, and the
H-rich layers had not been reached. Therefore only $\sim4.5$ pressure
scale heights needed to be included in those simulations which made them
considerably less demanding than the new simulations. In addition the
previous simulations were only in 2D.

The new simulations were performed on a cubical domain with two
uniform Cartesian grids of $576^3$ and $384^3$ respectively
(\fig{fig:3Dfluids}).\footnote{The $576^3$ calculation took 4 days on
  24 workstations at the University of Minnesota's Laboratory for
  Computational Science \& Engineering (LCSE). A movie made from the
  output of this run may be downloaded from the \htmlref{LCSE Web
    site}{http://www.lcse.umn.edu/index.php?c=movies}.}  Each 
simulation realistically represents the abundance mixture in the
He-shell flash convection zone and in the stable layer above as
different materials with the correct molecular weight ratio.  The
setup includes an inert white-dwarf-like core and a radiative region
below the bottom of the He-shell flash convection zone at
$9,500\mem{km}$ where the gravitational acceleration is
$4.9545\cdot10^7\mem{cm/s^2}$, the density is
$1.174\cdot10^4\mem{g/cm^3}$ and the pressure is
$1.696\cdot10^{20}\mem{g/cm\,s^2}$. At the bottom of the convection
zone a luminosity of $4.2\cdot10^7\lsun$ is artificially added in a
shell of $1,000\mem{km}$. This heating corresponds to the He-burning
that drives the flash, and compares as follows to the He-shell flash
luminosity in the stellar evolution models shown in
\fig{fig:SEprofiles}. In the model at time $t_\mem{0}$ the He-burning
luminosity is at its peak of $L_\mem{He,0} = 4.75\cdot10^7\lsun$
whereas it drops somewhat once the H-burning flash ignites at
$t_\mem{1}$ when $L_\mem{He,1} = 4.27\cdot10^7\lsun$. Thus, the 3D
hydrodynamic simulations are driven at the nominal heating rate.

The top of the convection zone is at a radius of $30,000\mem{km}$ and
surrounded by a radiative shell of thickness $4,500\mem{km}$. The
three layers are each polytropes. The adiabatic polytrope that
represents the convection zone spans $\sim9\mem{H_p}$. The setup
contains two materials. The lighter material represents the H/He
mixture in the stable layer above the convection zone. The heavier
fluid represents the \czw-rich mixture that occupies the convection
zone. We have assumed here that the material in the stable layer below
the convection zone has the same molecular weight. The ratio of the
molecular weights of the two components is
$\mu_\mem{C,O,He}/\mu_\mem{H,He} =2.26$.

\begin{figure}[tbp]
   \centering
   \includegraphics[width=0.6\textwidth]{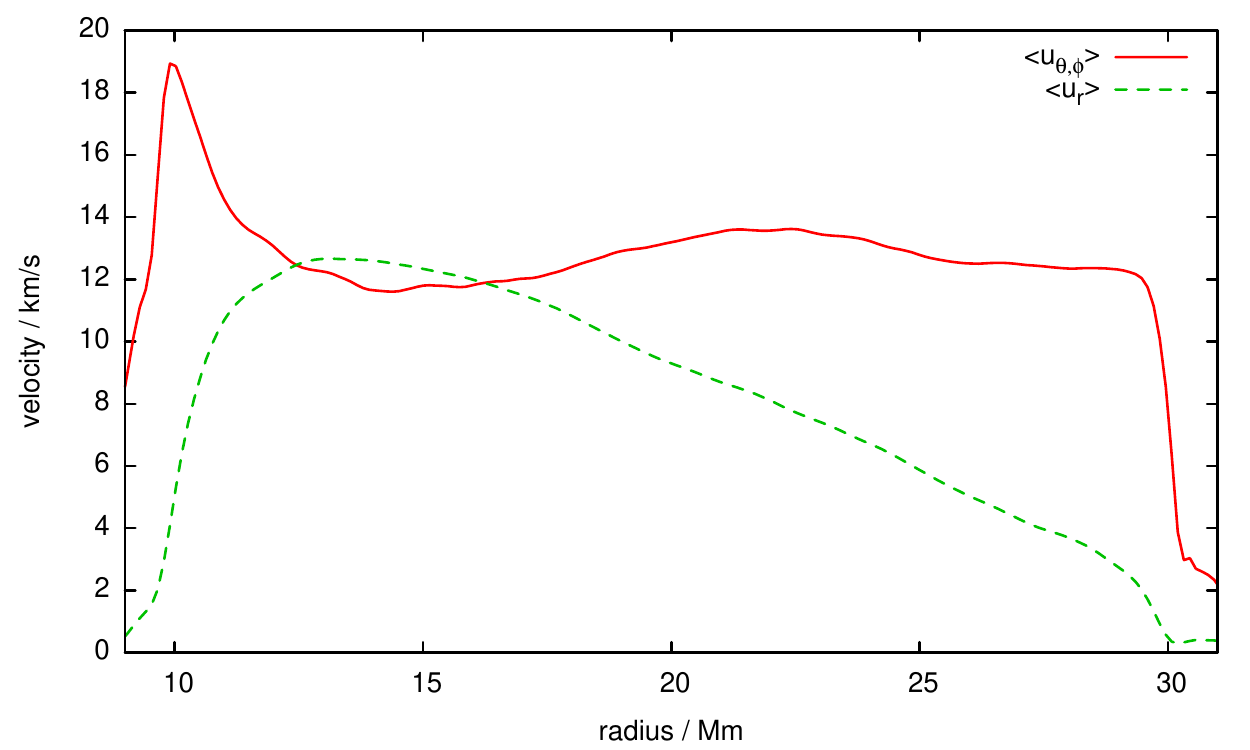} 
   \caption{Radial and tangential radially averaged rms-velocities of
     the $576^3$ simulation at the same time as shown (in the right
     panel) of \fig{fig:3Dfluids}. }
   \label{fig:v-profile}
\end{figure}

The higher resolution run (\fig{fig:3Dfluids}, right panel) is shown
at time $21,653\mem{s}$. For convective transport the typical radial
velocities are of interest. In the shown snapshot the largest radially
rms-velocities are found about $4,500\mem{km}$ above the bottom of the
convection zone around
$<v_\mem{rad,ave}>=\sqrt{2<E_\mem{kin}>}\sim12.5\mem{km/s}$. The
velocity of individual convective gusts can be significantly
higher. Towards the upper boundary of the convection zone the radial
velocities decrease to a few $\mem{km/s}$. This is compensated by
large tangential velocities $>12\mem{km/s}$ which stay this high all
the way to the convection boundary (\fig{fig:v-profile}). The
resulting strong radial gradient of the tangential velocities at the
top convection boundary is, via Kelvin-Helmholtz instabilities, likely
the main mechanism of the entrainment and convective boundary mixing
that we observe in these simulations. The information on typical
convective velocities together with the radial scale of the convection
zone implies a convective turn-over time scale of the order $\sim
3000\mem{s}$ (cf.\ Appendix \ref{sec:timescale}).  Therefore,
\fig{fig:3Dfluids} shows the entrainment after $\sim 7$ convective
turnovers.\footnote{We have continued this run for another 14
  convective turnovers. However, as will become clear from the
  following discussion the omission of proton burning limits the
  scientific use of that later part of the run to our
  application. Note that the time step of the 3D simulations is
  limited to $\Delta t = 5.9\cdot10^{-2}\mem{s}$ which implies that
  300,000 cycles had to be computed to reach the state shown.}  When
estimating the time scale for H-rich material to enter the convection
zone it must be considered that the entrained material is dominantly
transported in downflow lanes that are gravitationally compressed as
the material descends. This mechanism is reflected in the radial
velocities of the H-rich material that has entered the convection
zone, which in the snapshot shown exceed $20km/s$. We note that for
this component even the radially averaged velocity corresponds to a
Mach number $Ma \sim 0.02$ which is much higher than the MLT
convective velocity based estimate of $Ma\sim 0.001$.

\begin{figure}[tbp]
  \centering
   \includegraphics[width=0.7\textwidth]{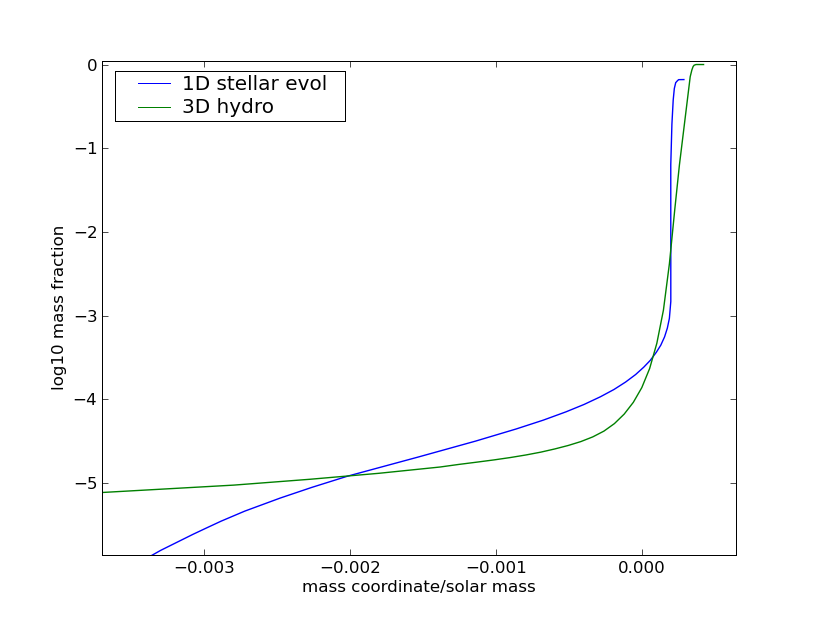} 
   \caption{Comparison of entrainment of material from the stable
     layer above the convection zone into the \czw-rich layer as it is
     represented in the one-dimensional stellar evolution model with mixing treated
     as diffusion in the mixing-length picture and in the 3D
     simulations discussed in this paper. The 3D profile (green line)
     shows the same data, radially averaged, as in \fig{fig:3Dfluids},
     right panel. The 1D line (blue) is the line labeled $t_\mem{0}$
     in \fig{fig:SEprofiles}. The mass coordinates have been set to
     zero in both cases near the top of the convection zone.  }
   \label{fig:1D3D-compare}
\end{figure}

After some initial transient period the convection assumes a flow
pattern that is dominated by large upwelling convective cells that
occupy typically a full octant as they emerge at the top convection
boundary. These large convective structures can be observed because we
simulate the full $4\pi$ sphere. Entrainment of the H-rich material
from the stable layer into the convection zone is mostly associated
with downdraft lanes that form when large cells collide on the surface
of the convection zone (\fig{fig:3Dfluids}). Note that the radially
averaged profile of the ingested H-rich material from the 3D
hydro-simulation is qualitatively very similar compared to the
diffusion picture of the one-dimensional stellar evolution
(\fig{fig:1D3D-compare}), at least close to the upper boundary. Futher
inward the lines divert from each other systematically as no H is
burned in the 3D simulations (this physics is not yet included).

However, the important result of the 3D simulations is that
entrainment is  rather inhomogeneous and asymmetric, as well as
intermittent in locally confined wedges of the star. From the
snapshot image of the entrainment it is clear that significant
anisotropy of the H-abundance is advected into the deeper layers where
the burning will eventually take place.

\subsection{Implications for the nucleosynthesis in a convective
  reactive environment like Sakurai's object}
\label{sec:hydro-implication}
We will give a full account of these simulations elsewhere. Here we
want to describe a few properties that are relevant for guiding our
mixing strategy for the nucleosynthesis simulation of the flash in
Sakurai's object.  The details of the convective-reactive burning of
hydrogen in the He-shell flash convection zone depend on two aspects
of the problem that hydrodynamic simulations can address. The
first is the process of entrainment. How much is the
fuel is premixed immediately after the entrainment in the
near-boundary layers. Subsequently these H-enriched fluid elements
will be carried along with the convective flow to deeper and hotter
layers where protons will eventually react with \czw.  This leads to the
second aspect of the problem, the hydrodynamic feedback of the nuclear
energy released. In the one-dimensional simulations this feedback is
in the form of a sharp entropy barrier, or a thin shell of positive
entropy gradient locally confined to a sphere. In reality the
thickness of this layer will depend on the velocity distribution and
the abundance distribution of fluid elements entering the layers hot
enough for rapid burning.

We can illustrate the possible outcomes by considering two extreme
cases. Assuming first that any entrained material is immediately
mixed and that vertical velocities of fluid elements are only
deviating negligibly from some average value (obviously, this case is
very close to the MLT picture of convection) then all fluid parcels or
blobs would release nuclear energy at almost the same radial position
inside the convection zone, and thus a very thin burn layer would
form, concentrating the entropy jump into a narrow region with large
positive entropy gradient, and soon inhibiting any further radial
mixing. The other extreme would be a wide range of mixing ratios in
blobs of H-enriched material entering the deeper layers with a large
range of velocities. Both of these inhomogeneities lead to a
broadening of the burning layer. To first approximation a blob (note
that this may be a shredded blob in order to conceptually overcome
mixing-length concepts) burns at $Da \sim 1$
(\kap{sec:intro_conv}). For smaller $Da$ (above the burning layer) the
nuclear reaction time scale is longer than the mixing time scale and
the blob will rather move further down than burn. For $Da>1$ we are
below the burning layer because now the blob burns faster than it can
move further down. Since the burn time scale decreases with depth a
range of mixing velocities translates into a spatial range in which
$Da\sim 1$.  Differently than in the first case, the velocity
distribution of blobs leads to a broadening of the burn
layer. Distributing the energy released from proton capture over a
thicker layer will make the emerging entropy gradient
shallower. Mixing accross the burn layer will be more efficient.  A
distribution of levels of H-enrichments in blobs being advected
through the burn layer would mean that the H-abundance is
heterogeneous (patchy) on spheres. Thus, the energy generation and the
dynamic feedback may very well be patchy and inhomogeneous on spheres,
as well as time variable.  At least initially, the inhibiting effect
of the burn layer on mixing may as well be time variable and
inhomogeneous on spheres.

In other words, an inhomogeneous distribution of fuel abundance in
blobs together with a distribution of vertical blob velocities would
have the tendency to delay the inhibiting effect of nuclear burning on
mixing from the top to the bottom of the convection zone. We leave a
detailed quantitative analysis of these processes to a forthcoming
investigation. Here we focus on the conceptual guidance we can gain
from the hydrodynamic simulations. These do indeed show a significant
inhomogeneity of the entrained material all the way down to the bottom
of the convection zone (\fig{fig:3Dfluids}), as well as a significant
distribution of vertical velocities, including convective gusts up to
Mach numbers around $Ma\sim0.03$.

We conclude from this analysis that the hydrodynamic nature of the
convective-reactive phase of H-ingestion into the He-shell flash
convection zone likely translates into a continued mixing through the
burn layer. We therefore hypothesize
that mixing is not inhibited at the early stage, as indicated by stellar
evolution models, but that instead mixing accross the H-burning layer is
possible for a prolonged period. It may stop only at a later time
after more H-ingestion has taken place. In the next section we will
test this hypothesis through nucleosynthesis simulations that
can be compared with the observations \citet{asplund:99a}.

\section{Nucleosynthesis simulations}
\label{sec:nuc2}

In this section we will describe mixing and nucleosynthesis
simulations based on the thermodynamic stellar evolution structure of
a post-AGB He-shell flash. We describe intially two cases, one that
resembles the mixing predicted by stellar evolution
(\sect{sec:sesimix}), and one with a mixing prescription that reflects
the findings discussed in the previous section
(\sect{sec:nucsimix}). While the first fails to reproduce key
observational features of Sakurai's object, the second one succeeds.
We show that high neutron densitities in the range $10^{12} <
N_\mem{n}/\mem{cm}^{-3} < 10^{16}$ are required to reproduce the
observed abundance features, as already pointed out by
\cite{asplund:99a}.  Such a neutron density regime is higher than the
classic $s$ process and significantly lower compared to the classic
$r$ process.

\subsection{General setup of nucleosynthesis simulations}
\label{sec:nucsetup}
We are using the MPPNP post-processing code (\sect{sec:mppnp}) to
calculate the nucleosynthesis of a He-shell flash peak one-dimensional
stellar structure model. We use two structures, one of them shown in
\fig{fig:SEprofiles} for $t=t_\mem{0}$. The MPPNP code reads the
mixing-length theory diffusion coefficient as well as the temperature
and density structure from the stellar evolution structure model. We
post-process this structure with sub-time steps of $\Delta
t_\mem{post-processing}=63\mem{s}$.  Thus, the mixing time scale is
well resolved, and the numerical splitting of the mixing and the
nucleosynthesis operators is justified.  The He-shell flash convection
zone is spatially resolved with $70$ to $90$ zones.  The grid is
statically refined and provides extra resolution near the ingestion
layer at the top of the convection zone, as well as around any split
region, should it be included.

The MLT based diffusion coefficient that is read in along with the
stellar structure from the stellar evolution output does not show a
split because the stellar evolution model is from a time just before
the ingestion of H-rich material begins. However, we are providing for
an optional split that can be inserted at an arbitrary location and
time, by modifying the diffusion coefficient in Eulerian coordinates
in the following way \beq D_\mem{with\ split} =
\frac{D_\mem{MLT}}{(1. + a_\mem{2}
  \exp(-a_\mem{1}(m_\mem{r}-m_\mem{r,split})^2)} \eeq where the split
is located at $m_\mem{r,split}$. When a split is imposed it is chosen
to be deep enough so that only very little material can be mixed
through, and the split is also very narrow. With $a_\mem{1}=10^4$ and
$a_\mem{2}=10^7$ the diffusion coefficient in the convection zone of
$D_\mem{MLT} \sim 5\cdot 10^{13}\mem{cm^2/s}$ is reduced to
$D_\mem{split,min} \sim 5\cdot10^{6}\mem{cm^2/s}$ over a width of $<
10^{-4}\msun$. We emphasise that $a_\mem{1}$ and $a_\mem{2}$ are free
parameters of our simple delayed split model and their particular
value is not important at this point. Only further hydrodynamic
simulations can possibly determine the mixing properties in this
environment. The purpose of the delayed split in terms of the radially
averaged nucleosynthesis calculations is further discussed below.

We are solving only for the nucleosynthesis and mixing equations while
the $T$, $\rho$ stratification is assumed to remain unchanged. Protons
and \hedr\ are inserted into the top of the convection zone at a rate
that is derived from the Lagrangian velocity of the top of the
convective boundary, as it moves into the H-rich layers above the
convection zone in the stellar evolution model. This velocity is
$\dot{M}_\mem{top,conv} \sim 1.7 \times 10^{-2}\msun/\jahre$. We are
ingesting at a rate of $5.3 \times
10^{-10}\msun/\mem{s}$.\footnote{Specifically, we add every $\sim
  6\mem{min}$ (every $6^\mem{th}$ cycle, corresponding roughly to 10
  times per convective turn-over time) $\Delta X = 5\times 10^{-4}$ to
  the mass fraction of H in the uppermost $4\times10^{-4}\msun$ of the
  convection zone. The baryon numbers are conserved by subtracting the
  required mass fraction from \czw. The abundances up to \nadr\ are
  initialized as described in \kap{sec:stellar_evol}.} We also add
\hedr\ according to the solar $\mem{H}/\hedr$ ratio in order to obtain
a prediction for Li.

Another constraint is that the total amount of H available for
ingestion is limited to the small remaining envelope mass that remains
on the pre-formed WD when the star leaves the AGB. For a core mass of
$0.6\msun$ this envelope mass is $\sim10^{-4}\msun$ with H and He
fractions as expected at the end of the AGB (mostly the initial ratio
possibly modified by third dredge-up). In all of the cases discussed
here we always find a nucleosynthetic reason to stop a simulation
before we run out of fuel.

\subsection{Stellar evolution mixing case}
\label{sec:sesimix}
\begin{figure}[tbp]
   \includegraphics[width=1.0\textwidth]{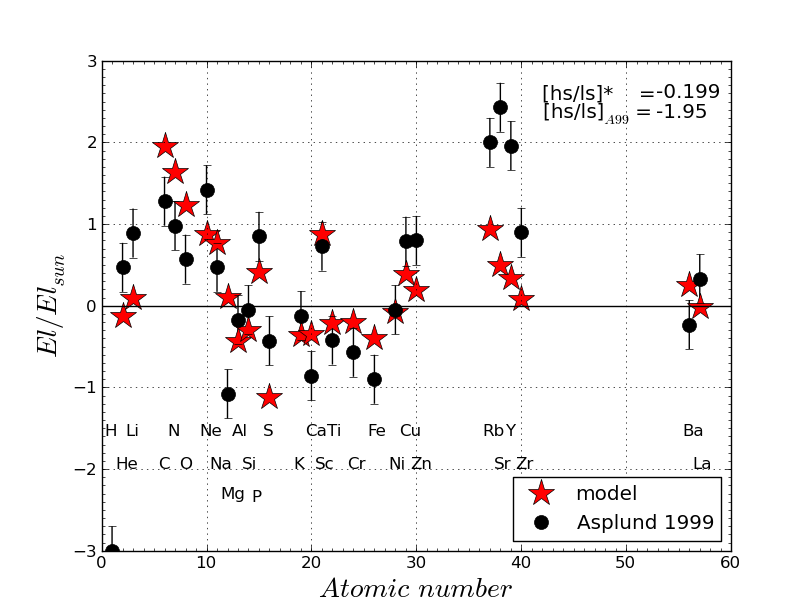}
   \caption{Abundance distribution obtained at the top of the He
     intershell assuming that the mixing split develops as soon as H
     is ingested. This case correspoonds to the one-dimensional
     stellar evolution prediction for mixing in the H-ingestion
     flash. The abundances measured by \cite{asplund:99a} are reported
     for comparison.  
}
\label{fig:ES}
\end{figure}

In the stellar evolution models the convection zone split due to
H-burning activation starts as soon as H is ingested (\kap{sec:nuc1}),
and no H or $^{13}$C can by mixed below the split coordinate.  In
Fig. \ref{fig:ES}, we show the abundance distribution prediction at
the top of the convection zone for this model in comparison with the
observations by \cite{asplund:99a}. We have used the $(\rho,T,D)$ stratification (strat-A)
from the \citep{herwig:99c} sequence, selecting a model just before
the H-ingestion starts as a template for this run. The mixing split as
described in the previous section is activated immediately as H starts
to mix into the convection zone. Peak H-burning is located at a
higher temperature in the \citep{herwig:99c} sequence  compared to more recent
models, and therefore this case yields an upper limit of the nucleosynthesis
efficiency predicted from one-dimensional models. \footnote{As
  discussed in \kap{sec:intro-sak} this older model did not reproduce
  the observed light curve, but more recent models
  \citet{herwig:01a,miller-bertolami:06} predict the split at lower
  temperature and as a result even less n-induced nucleosynthesis.}

 Calculations are run for about
one year, after which 
also Ba starts to be produced, in disagreement with observations.
 The neutron density reaches a value of
the order of 10$^{11}$ cm$^{-3}$ at the split coordinate due to the
high $^{13}$C concentration accumulated via proton capture on
$^{12}$C. This value is comparable with the neutron density obtained
at the bottom of a regular He-shell flash convection zone from
$^{22}$Ne($\alpha$,n)$^{25}$Mg reaction. Nevertheless, the predicted abundances do not matching the observations.

Li was produced initially during the ingestion \citep[see below
and][for more details]{herwig:00f} is destroyed on the time scale of
$\sim 1\jahre$. Stellar models predict that material around and beyond
the split expands and cools which reduces the $\alpha$-capture 
efficiency depleting Li.  But this also reduces the production of heavy elements.  

Sc is well reproduced within the uncertainties, in neutron densities
higher than in the classic $s$-process.  $^{40}$Ca is the main seed
along the neutron capture path, and Sc is mainly produced as $^{45}$Ca
which will decay to $^{45}$Sc in $\sim$ 166 days.  The production of
Sc is subject to nuclear reaction uncertainties, for instance
the (n,$\gamma$) rates of Ca isotopes, $^{41}$Ca(n,p)$^{41}$K and in
particular $^{41}$Ca(n,$\alpha$)$^{38}$Ar.

The bottom line is that Li observations cannot be reproduced together
with a significant $s$-process nucleosynthesis in this
simulation. But most importantly, the predicted [hs/ls] ratio much
higher than observed. Therefore, the nucleosynthesis simulation based
on the one-dimensional stellar evolution prediction for mixing cannot
account for the observed abundance patterns in Sakurai's object, which
confirms our findings from \sect{sec:nuc1}.

\subsection{Delayed split model motivated by the hydrodynamic
  simulations}
\label{sec:nucsimix}

We now assume that the split is not created instantaneously by
H-burning, but mixing continues --- at least initially ---
 unrestricted despite the energy generation from
H-burning (see \sect{sec:hydro-implication}). 
We use same background model (strat-A) as in \sect{sec:sesimix}.

\ndr\ is still formed in the upper layers where
the reaction and the mixing time coincide
(\abb{fig:profile}). \ndr\ decays to \cdr\ on a time scale of $\sim
10\mem{min}$. During this time \ndr\ will be swept along with the
flow, possibly covering a distance of the order
$10,000\mem{km}$. Eventually \cdr\ is mixed to the bottom of the
He-shell flash convection zone ($T\sim 2.5-3.0\cdot10^8\kelv$) and
establishes an abundance of $\sim 1\%$ by mass throughout the He
intershell.  Neutrons are released via $^{13}$C($\alpha$,n)$^{16}$O on
the time scale of $1\dots10\mem{s}$ and neutron densities reach a
value of $\sim 10^{15} \mem{cm}$ at the bottom of the convection zone.
The profile for Sr is shown as an example for how the abundance, even
of heavy elements, varies inside the convection zone as mixing and
production proceed at similar time scales.

\begin{figure}[tbp]
   \includegraphics[width=1.0\textwidth]{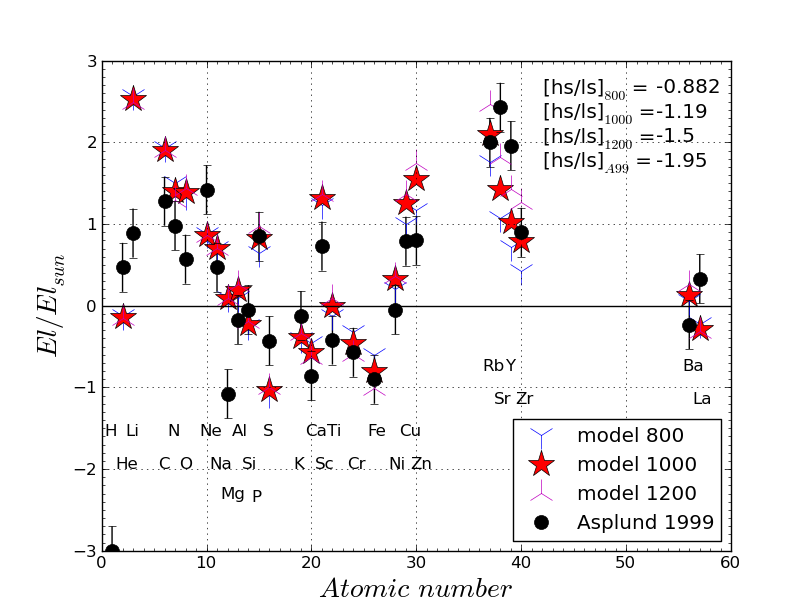}
   \caption{Abundance distribution at step 2000 for different cases
     with the split starting after $800\mem{min}$ (RUN105),
     $1000\mem{min}$ (RUN103) and $1200\mem{min}$ (RUN106).}
\label{fig:split}
\end{figure}

The intense neutron burst leads to the formation of the first
s-process peak elements Rb, Y, Sr, Zr, with Fe as the main seed.  The
unimpeded mixing between the formation region of \ndr\ and the deeper
layers where the neutrons are released must finish before the Ba-La
elements are significantly produced, which is not observed. This
defines the moment when mixing finally has to be limited, and we then
turn on the delayed split.  In Fig. \ref{fig:split} we show 
the abundances expected at the top of the He intershell
for different split time between $800\mem{min}$ to $1200\mem{min}$.

Burning of $^3$He produces $^7$Be via the reaction
$^3$He($\alpha$,$\gamma$)$^7$Be, which will decay later to $^7$Li.  As
pointed out by \citet{herwig:00f}, Li destruction is avoided under
these conditions not because Li is mixed into cooler regions
(Cameron-Fowler mechansim). Rather, in this \emph{hot H-deficient
  $^3$He-burning} all the protons are consumed before \besi\ decays to
\lisi. Then \lisi\ is more stable as it is only destroyed through
$\alpha$-captures.  In all cases Li is overproduced if we can assume
that a sufficient supply of \hedr\ is still available in the envelope
when the VLTP begins (cf.\ \kap{sec:intro-sak}).

Mg is more abundant in the simulations by one order of magnitude
compared to observations.  In all runs Mg is only weakly modified by
nucleosynthesis. For this reason, the low observed Mg abundance may be
another indicator of a sub-solar initial metallicity of the star,
unless there is some observational problem.

Despite the differences between these tests and the measurements, the
overall abundance trends are similar.  In particular all three test
cases in Fig \ref{fig:split} have a low [hs/ls] that ranges
between -0.9 and -1.5, decreasing with increasing the split delay.  In
the case with the latest split (at $1200\mem{min}$), [hs/ls] is still $\sim
0.5\mem{dex}$ higher than observed in Sakurai's object.  Light and
intermediate elements are not much affected by the split time.

\begin{figure}[tbp]
   \includegraphics[width=1.0\textwidth]{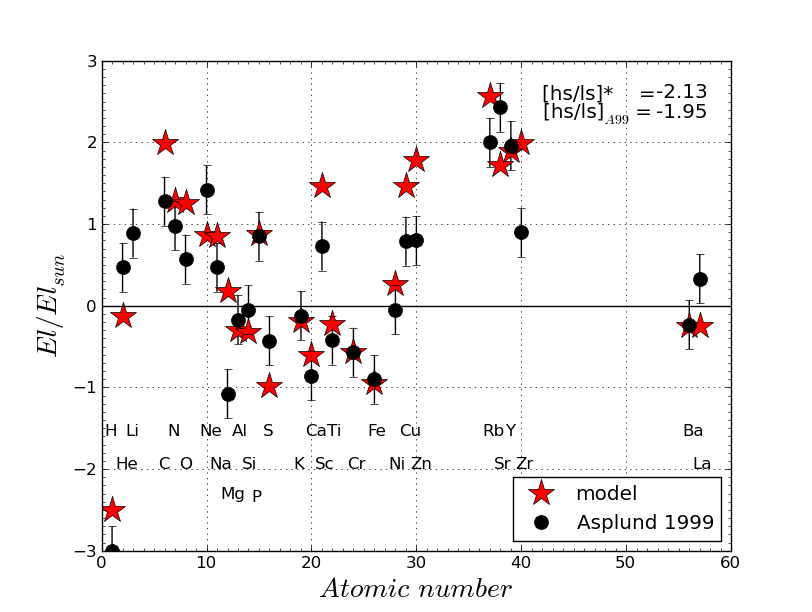}
   \caption{Abundance distribution at the end of the simulations (RUN48/strat-B)
     after $3000\mem{min}$, when all H- and \hedr-ingestion has been
     ingested.}   
\label{fig:13000}
\end{figure}

\begin{figure}[tbp]
   \includegraphics[width=1.0\textwidth]{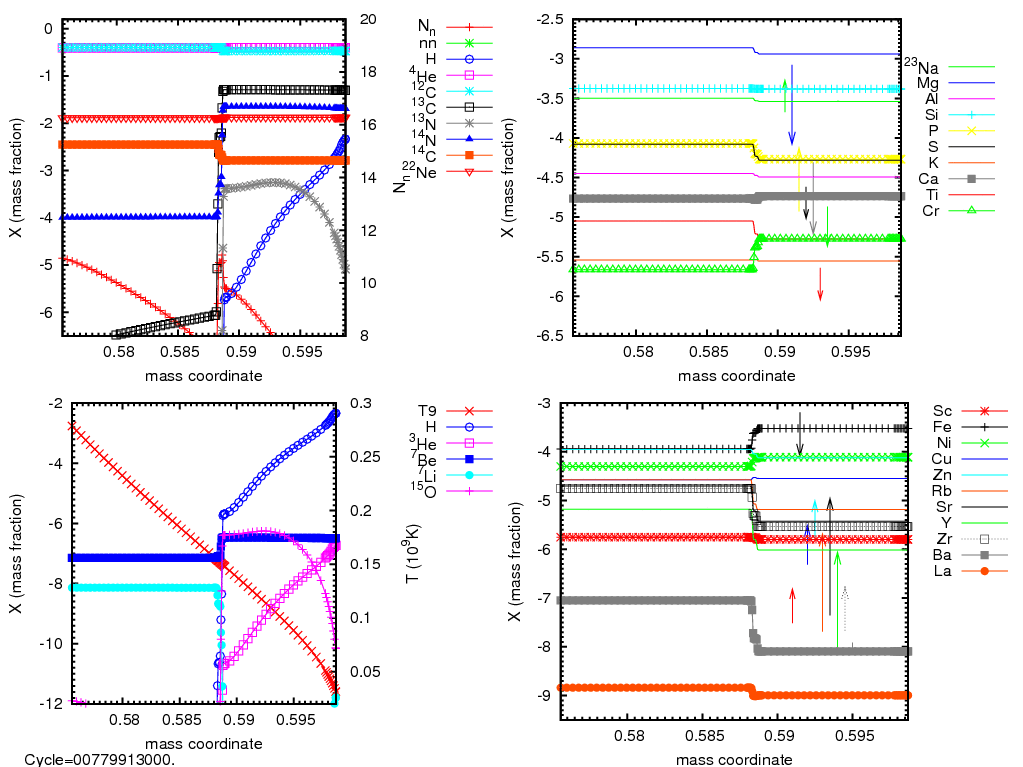}
   \caption{Abundance profile at the end of simulation RUN48 after
     $3000\mem{min}$, when all H- and \hedr\ has been ingested. A
     split imposed at $950\mem{min}$ has prevented further mixing
     between the He-shell flash driven convection zone (left) and the
     H-ingestion flash driven convection zone (right). Arrows in the
     right panels indicate the observed abundances, connecting the
     solar values with the observed ones.}
   \label{fig:pmun3000}
\end{figure}

In addition to the split delay time the quantitative model predictions
depend on the base stratification and convective mixing coefficient
taken from the stellar evolution model. This determines, for instance,
how quickly the protons and resulting \cdr\ are mixed, and in turn the
neutron density. To test the dependence of the results
on this point we present another set of simulations based on the
structure (strat-B) at the last thermal pulse in the $2\msun$ star
model sequence by \citet[model ET14][]{herwig:05a}. We have applied a
delayed split as for the strat-A model.  With this base structure, the
measured [hs/ls] is reproduced within the uncertainties
(Fig. \ref{fig:13000}). However, now Zr is higher by 1 dex compared to
the Asplund et al.\ measurements.  A general overview of the abundance
profiles 
in the He intershell for
the most indicative light isotopes and of the elements
included in Fig. \ref{fig:13000} is given in Fig. \ref{fig:pmun3000},
where the split position and the variation in the abundances are
shown.

Fig. \ref{fig:pmun3000} (left upper panel) confirms that the
$\czw/\cdr = 6.7$ ratio agrees within uncertainties with the observed
ratio of $\sim 3 \dots 5$.  $^7$Be is shown in the lower left panel
to be highly abundant, which will feed Li.

This profile view of one of our simulations reveals that the neutron
capture nucleosynthesis continues below the split, thereby further
modifying chemical abundances.  Possibly this further processed
material below the split has affected Sakurai's observed surface
abundances, through additional, later mixing.  H-burning at the split
must lose efficiency at some point when running out of fuel. This may
allow material exchange between the two regions \citep[see also
discussion in][]{asplund:99a}.  Asplund et al.\ observed Sakurai's
object four different times in 6 months, and these observations show
some drastic changes for some elements.  It is not the aim of this
paper to directly address these abundance trends over the sixth-month
period, since this level of detail cannot be captured by our modeling
approach, but has to await updated multi-dimensional simulations.

\begin{figure}[tbp]
   \includegraphics[width=1.0\textwidth]{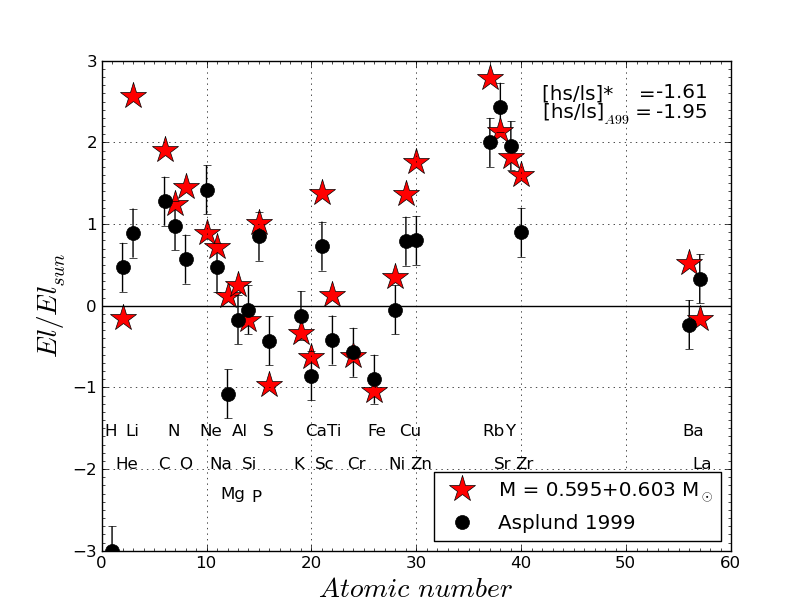}
   \caption{Abundance distribution  
     at step 2000 for a split delay of $1200\mem{min}$, considering
     mixing of $10\%$ of the deep component with
     $90\%$ from the component above the split.
   }   
\label{fig:mix}
\end{figure}

However, we may assume, as a working hypothesis, that the He intershell
is made of two components, one heavily processed below the split
(region 1), and one above the split (region 2) that was affected only
by the first ingestion phase. Because of the decreasing of efficiency of the H-burning
at the split, some material from region 1 is allowed to reach region 2
again and contribute to the observed abundance distribution.  Such
a two-component model is shown in \fig{fig:mix}. Starting from the
simulation based on stratification strat-A, with a delayed split after
$1200\mem{min}$ (see \fig{fig:split}) $10\%$ of the material is coming
from region 1, and $90\%$ from region 2.  No significant differences
are obtained compared to Fig. \ref{fig:split}.  However, this depends
on how much material is mixed from region 1 to region 2.  In 
this specific case,
such mixing implies a decrease on [hs/ls], but also an increase on Ba
production, not supported from the observations.  For this reason, at
present we cannot confirm or rule out such a double component
scenario.

\subsection{Nuclear reaction rate uncertainties}
In this nucleosynthesis scenario both H- and He-burning reactions, as
well as the n-capture reactions including those of short-lived
isotopes, are important. Especially, several elemental abundances, for
example Ti and Sc, are strongly dependent on s-process branchings
which requires extra accuracy from the nuclear physics data. As we
want to use this case to probe future hydroynamic simulations we need
to asses the influence of nuclear rate uncertainties.

\begin{figure}[tbp]
   \includegraphics[width=1.0\textwidth]{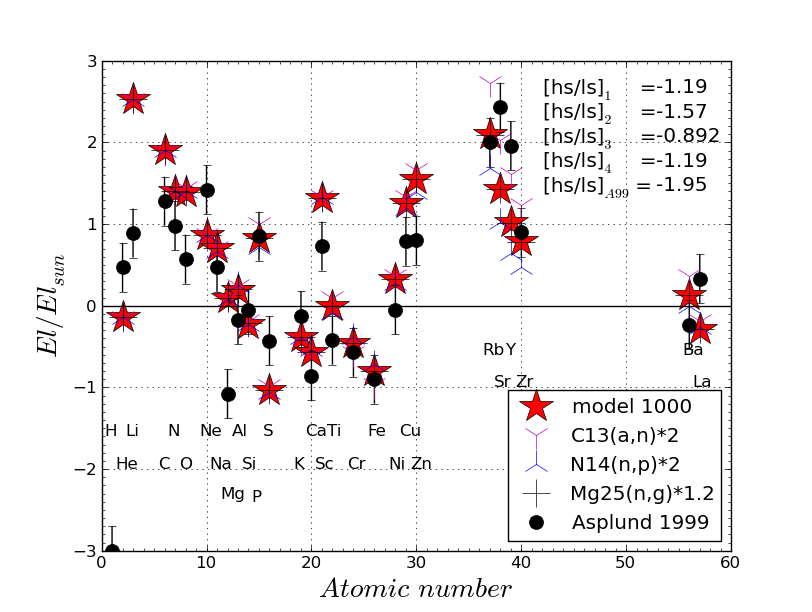}
   \caption{Abundance distribution for different nuclear test at step
     2000, from RUN103 (split delay $=1000\mem{min}$) as standard, and
     RUN107,RUN108,RUN109.  }
\label{fig:nuc}
\end{figure}

In Fig. \ref{fig:nuc} we show for the model strat-A with split after
$1000\mem{min}$ the impact of changing  the
$^{13}$C($\alpha$,n)$^{16}$O and the $^{14}$N(n,p)$^{14}$C reactions by a factor of $2$. The $^{25}$Mg(n,$\gamma$)$^{26}$Mg reaction has been varied  by a
factor of $1.2$. 
$^{13}$C($\alpha$,n)$^{16}$O is the main neutron source and the
two neutron capture reactions are important neutron poisons.   Among
these tests, the [hs/ls] changes between -0.9 and -1.6. In particular
the first peak elements are strongly affected. The Rb abundance changes by
$1 \mem{dex}$. Intermediate and light element predictions are only
weakly affected by nuclear reaction rate uncertainties.
Small errors associated with the CNO cycle rates 
(e.g., $^{12}$C(p,$\gamma$)$^{13}$N and $^{14}$N(p,$\gamma$)$^{15}$O)
have a marginal impact in our results compared to the other rates that we
have considered.

In Fig. \ref{fig:nuc}, we only included the impact of varying the 
neutron capture reaction rates of light neutron poisons.
In the short time scale of the neutron burst, the neutron capture process 
is also expected to show a strong propagation effect in the final abundance 
distribution beyond iron, due to uncertainties of neutron capture rates 
along the nucleosynthesis path.
In particular, such propagation may be relevant in our case, 
since Rb, Sr, Y and Zr production is affected by the error of several 
low cross sections of isotopes in the mass region between Fe and Sr, 
acting like bottle-necks in the neutron capture flow \citep[e.g., $^{62}$Ni,
$^{68}$Zn, $^{74}$Ge and $^{78}$Se][and reference therein]{pignatari:10}.
Another point to consider is that in the high neutron density regime
reached in our calculations several unstable isotopes are produced efficiently,
and many stable isotopes receive a significant contribution from unstable species
from radiogenic decay and/or from decay during the neutron freezout, 
when the split is established.
For instance, in all the cases presented in Fig. \ref{fig:nuc} most of Y (that is formed by
one stable isotope only, $^{89}$Y) is produced as $^{89}$Sr.
The neutron capture rates of unstable species are mostly theoretical, and 
also their large uncertainty (typically a factor of 2-3) may affect the final 
isotopic distribution.

None of our simulations seem to be reproducing Sc particularly well.
Sc and the elemental ratio Sc/Ca are particularly sensitive to the
neutron density. Indeed, $^{45}$Sc is produced as unstable $^{45}$Ca
via neutron captures on stable Ca species, where $^{40}$Ca is the main
seed for Sc production.  $^{41}$Ca is unstable, and has stronger (n,p)
and (n,$\alpha$) than (n,$\gamma$) channels.  For this reason, the
uncertainty in the relative efficiency of the (n,p), (n,$\alpha$) and
(n,$\gamma$) channels may affect the total Sc production.  Among
nuclear uncertainties, another possible explanation for Sc
overproduction is that the initial metallicity of the Sakurai's object
is even lower than what we have used for our simulations
([Fe/H]=$-$0.18).  Indeed, a lower initial $^{40}$Ca will results in a
lower final Sc abundance.

\section{Conclusions}
\label{sec:disc}

\subsection{Summary}
We have presented in this paper a multi-physics view of the combustion
in a very-late thermal pulse in a pre-WD. H is mixed convectively into
the He-shell flash convection zone. We have discussed the
one-dimensional stellar evolution picture, that predicts that early on
the energy generation from the $\czw(\p,\gamma)^{13}$N reaction creates a
sharp  entropy discontinuity which prohibits mixing. A detailed nucleosynthesis
analysis, based on a complete multi-zone treatment of nucleosynthesis with
mixing, shows that this one-dimensional structure evolution leads to
abundance predictions that are incompatible with the observed
abundances in Sakurai's object. Seeking guidance from full three-dimensional
hydrodynamic simulations of He-shell flash convection in $4\pi$
geometry with entrainment, we obtain reasons to suspect that the
burning front is more distributed than predicted in one-dimensional
stellar evolution. Fuel will be transported down in down-draft lanes
leading to an inhomogeneous distribution of fuel in the burning
zone. In addition, vertical down drafts enriched with fuel will populate a
velocity distribution. From this information we speculate that mixing
of protons and of the neutron source material \ndr\, which later becomes \cdr, across the
convective H-burning zone will proceed for much longer than indicated
by one-dimensional stellar evolution. 

We point out that the main nucleosynthetic signature of
convective-reactive burning in this study is the significant
overproduction of the first peak elements Sr, Y and Zr, coupled with a
non-efficient production of the second peak elements Ba and La.
According to our simulations, neutron densities 10$^{12}$ cm$^{-3}$ <
N$_{\rm n}$ < 10$^{16}$ cm$^{-3}$ are required to explain such
abundance distribution.  More specifically, in the Sakurai's object
time scale of $\sim$ 2 years between the luminosity peak due to H
burning and the Asplund's observations, a neutron density peak of
$\sim$ 10$^{15}$ cm$^{-3}$ with a delay of $\sim$ 1 day before the
complete split activation would qualitatively reproduce the observed
[hs/ls] ratio, the Li abundance and the low $^{12}$C/$^{13}$C
ratio.  The problems that we encounter in reproducing single
elements may be due to the approximations in our model (e.g., for the
nucleosynthesis simulations we use parameters from one-dimensional
stellar models), to observation problems (e.g., the observed Y/Zr
ratio cannot be reproduced by neutron capture nucleosynthesis) or to
nuclear physics uncertainties (e.g., Sc).

Nuclear reaction rate uncertainties are shown to have a
particularly important effect on some key observables in this
non-equilibrium nuclear burning environment.

\subsection{Implications for stellar evolution and nucleosynthesis of
  the first generations of stars}

One of our main motivations to study convective-reactive phases in
stellar evolution is their prevalence in models of the
first generation of stars. As reviewed in \sect{sec:intro_conv},
convective mixing of protons with the \czw\ from He-burning at
He-burning temperatures is frequently encountered in stellar evolution
calculations at very low and zero metal content at all masses. This
investigation shows that the predictive power of one-dimensional
stellar evolution simulations is severely limited for observables that
depend on these convective-reactive phases.

Neutron burst nucleosynthesis of the type described in this paper are
nevertheless expected to happen also in the convective-reactive H-\czw\
combustion events in first generation of stars.  Indeed, the neutron
source $^{13}$C is of primary origin, i.e. its abundance is not
affected by the metal content in the initial stellar composition.
Massive stars at different metallicities may experience H-\czw\
combustion, ingesting protons in the He shell \citep[see discussion
in][]{woosley:95}. If enough hydrogen fuel is ingested then Sr, Y and
Zr may be efficiently produced by the primary
$^{13}$C($\alpha$,n)$^{16}$O neutron source, just as in our
simulations presented here. This may be an alternative or
complementary explanation for a missing component in the first
neutron-peak region of the abundance distribution in both the solar
abundance distribution as well as the metal poor stars,
\citep[light-element primary process, or
LEPP ][]{travaglio:04,pignatari:08a,farouqi:09}. In the future we
intend to study the speculation that the convective-reactive
proton-\czw\ combustion in the convective He shell in massive stars
could provide another possible solution for the LEPP.

\acknowledgments FH acknowledges NSERC Discovery Grant funding. The
hydrodynamics simulations were performed by PRW on a cluster of
workstations at the Univesity of Minnesota, provided through an NSF
equipment grant, NSF-CNS-0708822. The work of CF and GR was funded in part
under the auspices of the National Nuclear Security Administration of
the U.S. Department of Energy at Los Alamos National Laboratory and
supported by Contract No. DE-AC52-06NA25396. RH acknowledges support
from the World Premier International Research Center Initiative (WPI
Initiative), MEXT, Japan. This work used the SE library (LA-CC-08-057)
developed at Los Alamos National Laboratory as part of the NuGrid
collaboration; SE makes use of the HDF5 library, which was developed
by The HDF Group and by the National Center for Supercomputing
Applications at the University of Illinois at Urbana-Champaign.


 \clearpage



\begin{table*}                      
\begin{center}                                      
\caption{\label{tab:asplund_abundances} Observed neutron capture signature, \cite{asplund:99a}.}
\begin{tabular}{c|rr}                                        
\hline 
\hline
 $[Fe/H]$ = 0 (-0.63)    &  April 1996   & October 1996 \\
\hline 
\hline
 $[Y/Fe]$   &  +0.96 (+1.59) &  +1.96 (+2.59)  \\
\hline
 $[Ba/Fe]$  &  -0.63 (0.0)   &  -0.23 (+0.40)  \\
\hline
 $[Ba/Y]$   &  -1.59 (-1.59) &  -2.19 (-2.19)  \\
\noalign{\smallskip}
\hline
\end{tabular}
\label{hsls:sakurai}
\end{center}
\end{table*}



\begin{figure}[tbp]
   \includegraphics[width=1.0\textwidth]{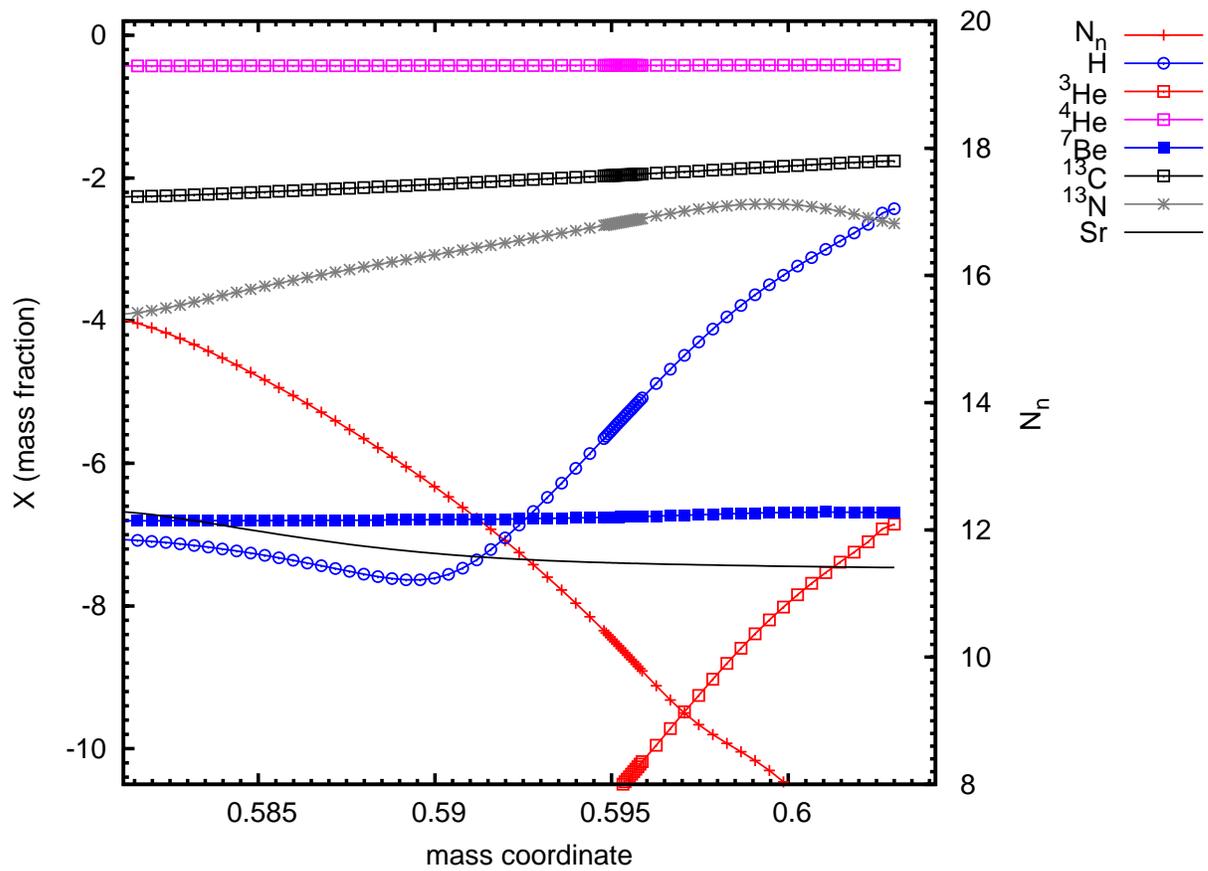}
   \caption{The abundance profiles snapshot (RUN103), just before the
     mixing split is imposed, demonstrates the simultaneous action of
     nucleosynthesis and mixing on similar time scales.}
\label{fig:profile}
\end{figure}

\clearpage

\appendix
\section{Code description}
\label{sec:codes}
\subsection{Nucleosynthesis}
\label{sec:mppnp}
The PPN physics package allows a flexible combination of nuclear
reaction rates and entire compilations of rates. For this study we
choose for the main charged particle reactions the compilation by
\citet{angulo:99} (NACRE compilation).
This choice allows us to be consistent with the original network
used to calculate the stellar structures for basic energetic nuclear
reactions, i.e., $^{14}$N(p,$\gamma$)$^{15}$O, 3-$\alpha$ and 
$^{12}$C($\alpha$,$\gamma$)$^{16}$O.
Notice that the use of more recent rates 
\citep[e.g.,][respectively]{imbriani:05,fynbo:05,kunz:02}
would not change our results, where uncertainties related to
physics processes and mixing still has a critical impact.
Other charged particle reactions, among the others 
$^{13}$C($\alpha$,n)$^{16}$O, which is the main neutron source during the 
H ingestion, have more recent measurements \citep[e.g.,][]{heil:08b}.
However, in this case NACRE rates are consistent with the new rates
within their uncertainty. For instance, we consider a factor of two of 
uncertainty for the $^{13}$C($\alpha$,n)$^{16}$O rate in the temperature
regime that is relevant for the $^{13}$C burning 
(see Section \ref{sec:nuc2} for more details).
For neutron capture reactions of stable isotopes we refer to \citet{dillmann:06}
(KADoNIS compilation).
Stellar $\beta$-decay rates and electron
captures are from \citet{oda:94} and \citet{fuller:85} for many light
unstable isotopes, and \citet{goriely:99} for many heavy unstable
isotopes. Rates not included in the previous references are given by
the Basel REACLIB compilation. We are solving the complete network in
each radial grid point, including all relevant charged particle,
n-capture reactions as well as the $\beta$-decays. A recursive,
dynamic network generation has been integrated into the solver, i.e.\
the size of the network automatically adapts to the conditions
given. If, for example, a neutron source is activated the network will
be automatically enlarged to include all heavy and unstable isotopes
as needed according to the network fluxes. This dynamic network
feature ensures that the network calculation never misses any
important isotope or reaction.

In these simulations we are using the multi-zone driver of the PPN code (MPPNP) that
allows for the calculation of the complete nucleosynthesis in all of
the zones of one-dimensional profiles, e.g.\ from stellar evolution,
of density and temperature. The MPPNP driver employs MPI parallelism
to enable efficient calculations on up to 30-50 processors depending
on problem sizes. The simulations carried out here involve relatively
small grids between 70 and 90 zones. A fully implicit nucleosynthesis step is
followed by a mixing step according to the diffusion coefficient
taken, for example, from the stellar evolution model. This procedure
is repeated for subsequent time steps in order to compute the
evolution of the abundance profiles of all species involved. Mixing
and network calculations are performed in the operator split mode,
which is a good approximation for the post-processing because we
choose the post-processing time step to be small enough to resolve the
mixing time scale.

\subsection{Hydrodynamics}
\label{sec:code-hydro}
 The PPM gas dynamics
scheme \citep{woodward:84,colella:84,woodward:86,woodward:06b} has
been in use in computational astrophysics for many years.  It is
incorporated in the community codes VH1 \citep{blondin:93}, ENZO
\citep{bryan:95}, and FLASH \citep{calder:02}.  The version that we
use in this work is described in full in \citet{woodward:06b}.  Here
we have augmented PPM with the PPB moment-conserving advection scheme
to treat the entrainment of fluid from above the convection zone
during the helium shell flash in an AGB star
\citep[see][]{woodward:08a}\nocite{woodward:08b}.  PPB is built upon van
Leer's Scheme VI \citep{vanLeer:77}, a 1-D scheme that conserves the
first 3 moments of the advected distribution in each grid cell.  To
this scheme we have added a set of very carefully constructed
constraints \citep{woodward:05} keeping the advected fractional volume
of a multifluid constituent of the gas within the range from 0 to 1.
These constraints are a considerable improvement over those outlined
in Woodward 1986 for a 2-D PPB scheme.  We have also streamlined the
implementation of PPB in 3-D by eliminating various high-order terms
in order to obtain a highly efficient, directionally split scheme
\citep{woodward:05} that conserves 10 moments of the distribution of
the advected fractional volume variable in each cell.  PPB is combined
with PPM to describe multifluid hydrodynamics by adding the constraint
of pressure and temperature equilibrium within each grid cell.  At
present our code is explicit.  Mach numbers in the convective gusts of
helium shell flash convection are about 1/30 or less.  Consequently,
we must take many time steps to follow the flow through an entire
circuit of a large convective eddy.  We note that such eddies are
global in scale, and we follow them by including the entire convection
shell in our computational domain.  The conclusion that large scales
are involved here is similar to the earlier findings of
\citet{porter:00b} and \citet{porter:06} for the outer convective
envelope of an AGB star.  The restricted time step values, from
explicit hydrodynamics, and the large domain, arising from the natural
scale of the convection, place significant demands on the computation.
We address these demands in two ways.  First, we exploit a new
implementation of our codes aimed specifically at the multicore CPUs
found in modern computers \citep[see][]{woodward:08c,woodward:09},
which has delivered to our codes roughly a 40x speed-up over
performance on single-core platforms from about 4 years ago (the code
performance now stands at 24 Gflop/s/4-core-CPU, scalable to thousands
of CPUs, and we obtain sustained performance over 1 Tflop/s on our
small local cluster daily).  Second, we exploit the fact that explicit
computation is roughly as efficient as implicit computation when Mach
numbers are around 1/30.

 The code scales to hundreds of thousands of
processor cores, for which runs with the proper heating rates, the
full convection zone, and well resolved entrainment at the convection
zone boundary are easily carried out in a single day.

\section{Time and Length Scales}
\label{sec:timescale}

The relevant nuclear burning time scale for the H-ingestion problem is
the time scale for a proton to be captured by a \czw:
\[ \tau_{^{12}\mem{C}}(\p) = \frac{12}{X(^{12}\mem{C})\, \rho \, \mathrm{N_a}<\sigma
  v>_{^{12}\mem{C}(\p,\gamma)}} \punkt\] 
For the quantitative evaluation of the relevant time scales we use the
pre-ingestion model at time $t_\mem{0}$ shown in
\abb{fig:SEprofiles} (\sect{sec:nuc1}).  The mass fraction of \czw\ in
that model 
is $X(^{12}\mem{C})=0.36$ and the density
increases from $\rho_\mathrm{top} = 1.26\times10^2\mathrm{g /cm^3}$
at the top of the convection zone to $\rho_\mathrm{bot} =
1.0410\times10^4\mathrm{g /cm^3}$ at the bottom of the convection
zone. The nuclear reaction rate $<\sigma v>_{^{12}\mem{C}(\p,\gamma)}$
depends sensitively on the temperature which increases from
$T_\mathrm{top}=2.2\times10^7\kelv$ at the top to
$T_\mathrm{bot}=2.9\times10^8\kelv$ at the bottom of the convection
zone. $ <\sigma v>_{^{12}\mem{C}(\p,\gamma)}$ increases by 12 orders
of magnitude accross the convection zone. 

The location of the peak H-burning due to H-ingestion takes place
where the mixing time scale is the same as $\tau_{^{12}\mem{C}}(\p)$
\citep[Ch.\,4][]{1996snih.book.....A}. The diffusion coefficient
$D_\mem{MLT}$ for convective mixing is derived from the mixing-length
theory (MLT). With an appropriate length scale $l$ a mixing time scale
can be obtained. For some properties the MLT mixing-length
$l_\mem{MLT}$ should be used: $l_\mem{MLT}=\alpha_\mem{MLT}H_\mem{p}$
with $\alpha_\mem{MLT}=1.7$ the mixing-length paramter and $H_\mem{p}$
the pressure scale height. This MLT mixing time scale is then
$\tau_\mem{MLT} =l_\mem{MLT}^2/D_\mem{MLT}$.  As can be seen in
\abb{fig:tau} $\tau_{^{12}\mem{C}}(\p)=\tau_\mem{MLT}$ at
$m_\mem{r}=0.6024\msun$, a significantly larger mass coordinate than
the location of the peak H-burning ($\sim0.6005\msun$) calculated in
the stellar evolution model, as evident from the H-profile at
$t_\mem{1}$ in \abb{fig:SEprofiles}.

$l_\mem{MLT}$ should not be used to estimate a mixing time scale
relevant for rapid nuclear burning, since the rate of p-captures  depends
only indirectly on $P$.  In fact, in the vicinity of the H-peak
luminosity the pressure scale height is $H_\mem{P}\sim 1.4\mem{Mm}$
which implies $l_\mem{MLT}\sim2.4\mem{Mm}$. This is much larger than
the distance over which the rate of p-capture by \czw\ (rate of
reaction) increases significantly. It is this rate of reaction length
scale that defines the width and location of the combustion flame for a
given diffusion coefficient. A generalized length for any quantity
$\phi=\phi(r)$ may be defined as \citep{chapman:61}
\[H_\mem{\phi}=\frac{1}{\frac{d\ln \phi}{dr}}  \] 
where $H_\mem{\phi}$ is the rate of reaction length
scale if we define $\phi = \rho\,  \mathrm{N_a}<\sigma v>_{^{12}\mem{C}(\p,\gamma)}$.
The rate of reaction mixing time scale is then
$\tau_\mem{\phi}=H_\mem{\phi}^2/D_\mem{MLT}$. As shown in
\abb{fig:tau} the mass coordinate where $\tau_{^{12}\mem{C}}(\p) =
\tau_\mem{\phi}$ coincides very well with the location of peak
H-burning (where as a result the mixing split 
occurs) at $t=t_\mathrm{1}$ in \abb{fig:SEprofiles}.

At this location ($m_\mem{r}\sim0.6005\msun$) the reaction length scale is
$H_\mem{\phi}\sim 330\mem{km}$ which is the geometric scale of the
flame that hydrodynamic simulations including nuclear burn have to
resolve. A simulation box that fits the $4\pi$ geometry of the entire
He-shell flash convection zone needs to have a side length of
$50\mem{Mm}$ which corresponds to $\sim166$ flame widths. In order to
resolve the flame with at least 10 radial zones an aequidistant grid
for a H-ingestion flash simulation needs to have a $1660^3$ grid.

\begin{figure}[tbp]
  \includegraphics[width=0.8\textwidth]{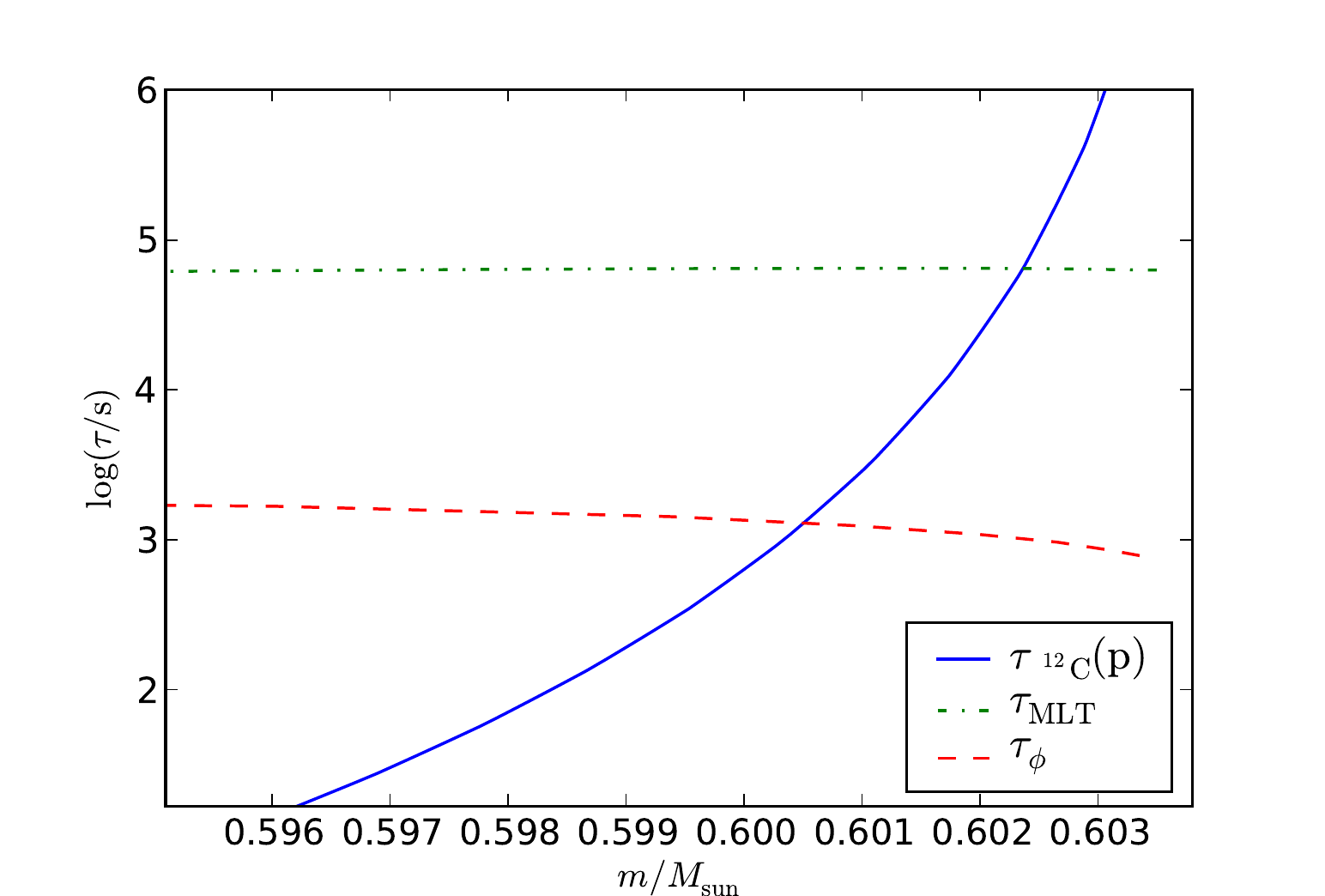}
  \caption{Time scales as a function of the mass coordinate in the
    convection zone, for proton capture by \czw\ (blue solid line), as
    well as the MLT mixing time scale (green dash-dot) and the
    rate of reaction mixing time scale (red dashed) (see text for
    details). For this figure the tabulated reaction rate from
    \citep{angulo:99} was used.  }
\label{fig:tau}
\end{figure}

\end{document}